\documentclass[fleqn,usenatbib]{mnras}

\usepackage{graphicx}
\usepackage{txfonts}
\usepackage{xcolor}
\usepackage[T1]{fontenc}

\def\lesssim{\mathrel{\hbox{\rlap{\hbox{\lower4pt\hbox{$\sim$}}}\hbox{$<$}}}}
\def\gtrsim{\mathrel{\hbox{\rlap{\hbox{\lower4pt\hbox{$\sim$}}}\hbox{$>$}}}}
\newcommand{\gps}{\ensuremath{g_{\rm P1}}}
\newcommand{\rps}{\ensuremath{r_{\rm P1}}}
\newcommand{\ips}{\ensuremath{i_{\rm P1}}}
\newcommand{\zps}{\ensuremath{z_{\rm P1}}}

\newcommand{\wps}{\ensuremath{w_{\rm P1}}}
\newcommand{\grizy}{\ensuremath{grizy_{\rm P1}}}

\newcommand{\degree}{\mbox{$^\circ$}}
\newcommand{\msun}{\,\mbox{M$_{\odot}$}}
\newcommand{\msol}{\,\mbox{M$_{\odot}$}}

\newcommand{\rsun}{\mbox{R$_{\odot}$}}
\newcommand{\kms}{\mbox{\,$\rm{km}\,s^{-1}$}}
\newcommand{\logl}{\mbox{$\log L/{\rm L_{\odot}}$}}

\newcommand{\CaII}{Ca\,{\sc ii}}

\newcommand{\NaI}{Na\,{\sc i}}

\newcommand{\cmg}{\,$\mathrm{cm}^{2}\,\mathrm{g}^{-1}$}
\newcommand{\kmsMpc}{\,$\mathrm{km\,s}^{-1}\,\mathrm{Mpc}^{-1}$}

\newcommand{\kpc}{\,kpc}
\newcommand{\Mpc}{\,Mpc}
\newcommand{\magday}{\,mag\,day$^{\mathrm{-1}}$}

\title[PS15cey and PS17cke]{PS15cey and PS17cke: prospective candidates from the Pan-STARRS Search for Kilonovae}

\author[O. R. McBrien et al.]{O. R. McBrien$^{1}$\thanks{E-mail: omcbrien02@qub.ac.uk (OMB)}, 
S. J. Smartt$^{1}$,
M. E. Huber$^{2}$,
A. Rest$^{3}$,
K. C. Chambers$^{2}$,
\newauthor
C. Barbieri$^{4,5,6}$,
M. Bulla$^{7}$, 
S. Jha$^{8}$,  
M. Gromadzki$^{9}$,
S. Srivastav$^{1}$,
K. W. Smith$^{1}$,
\newauthor
D. R. Young$^{1}$,
S. McLaughlin$^{1}$,
C. Inserra$^{10}$, 
M. Nicholl$^{11,12}$,
M. Fraser$^{13}$, 
K. Maguire$^{14}$, 
\newauthor
T.-W. Chen$^{15}$,
T. Wevers$^{16,17}$,
J. P. Anderson$^{17}$,
T. E. M\"uller-Bravo$^{18}$,
F. Olivares E.$^{19}$,
\newauthor
E. Kankare$^{20}$,
A. Gal-Yam$^{21}$,
C. Waters$^{22}$
\\
$^{1}$Astrophysics Research Centre, School of Mathematics and Physics, Queen’s University Belfast, BT7 1NN, UK\\
$^{2}$Institute for Astronmy, University of Hawaii, 2680 Woodlawn Drive, Honolulu, Hawaii 96822, USA\\
$^{3}$Space Telescope Science Institute, 3700 San Martin Drive, Baltimore, MD 21218, USA\\
$^{4}$Università degli Studi di Milano-Bicocca, Dipartimento di Fisica ``G. Occhialini'', Piazza della Scienza 3, I-20126 Milano, Italy\\
$^{5}$INAF -- Osservatorio Astronomico di Brera, via E. Bianchi 46, I-23807 Merate, Italy\\
$^{6}$INFN -- Sezione di Milano-Bicocca, Piazza della Scienza 3, I-20126 Milano, Italy\\
$^{7}$Nordita, KTH Royal Institute of Technology and Stockholm University, Roslagstullsbacken 23, SE-106 91 Stockholm, Sweden\\
$^{8}$Department of Physics and Astronomy, Rutgers the State University of New Jersey, 136 Frelinghuysen Road, Piscataway, New Jersey 08854, USA\\
$^{9}$Astronomical Observatory, University of Warsaw, Al. Ujazdowskie 4, 00-478 Warszawa, Poland\\
$^{10}$School of Physics \& Astronomy, Cardiff University, Queens Buildings, The Parade, Cardiff CF24 3AA, UK\\
$^{11}$Birmingham Institute for Gravitational Wave Astronomy and School of Physics and Astronomy, University of Birmingham, Birmingham B15 2TT, UK\\
$^{12}$Institute for Astronomy, University of Edinburgh, Royal Observatory, Blackford Hill, EH9 3HJ, UK\\
$^{13}$School of Physics, University College Dublin, Belfield, Dublin 4, Ireland\\
$^{14}$School of Physics, Trinity College Dublin, The University of Dublin, Dublin 2, Ireland\\
$^{15}$The Oskar Klein Centre, Department of Astronomy, Stockholm University, AlbaNova, SE-10691 Stockholm, Sweden\\
$^{16}$Institute of Astronomy, University of Cambridge, Madingley Road, CB3 0HA, UK\\
$^{17}$European Southern Observatory, Alonso de C\'ordova 3107, Casilla 19, Santiago, Chile\\
$^{18}$School of Physics and Astronomy, University of Southampton, Southampton, Hampshire, SO17 1BJ, UK\\
$^{19}$Instituto de Astronom\'{\i}a y Ciencias Planetarias, Universidad de Atacama, Copayapu 485, Copiap\'o, Chile\\
$^{20}$Department of Physics and Astronomy, University of Turku, Vesilinnantie 5, FI-20014 Turku, Finland\\
$^{21}$Department of Particle Physics and Astrophysics, Weizmann Institute of Science, Rehovot 76100, Israel\\
$^{22}$Department of Astrophysical Sciences, 4 Ivy Lane, Princeton University, Princeton, NJ 08544\\
}

\date{Accepted XXX. Received YYY; in original form ZZZ}

\pubyear{2020}

\begin{document}
\maketitle

\begin{abstract}
Time domain astronomy was revolutionised with the discovery of the first kilonova, AT2017gfo, in August 2017 which was associated with the gravitational wave signal GW170817.
Since this event, numerous wide-field surveys have been optimising search strategies to maximise their efficiency of detecting these fast and faint transients. 
With the Panoramic Survey Telescope and Rapid Response System (Pan-STARRS), we have been conducting a volume limited survey for intrinsically faint and fast fading events to a distance of $D\simeq200$\,Mpc.
Two promising candidates have been identified from this archival search, with sparse data - PS15cey and PS17cke.
Here we present more detailed analysis and discussion of their nature.
We observe that PS15cey was a luminous, fast declining transient at 320\Mpc.
Models of BH-NS mergers with a very stiff equation of state could possibly reproduce the luminosity and decline but the physical parameters are extreme.
A more likely scenario is that this was an AT2018kzr-like merger event.
PS17cke was a faint and fast declining event at 15\Mpc.
We explore several explosion scenarios of this transient including models of it as a NS-NS and BH-NS merger, the outburst of a massive luminous star, and compare it against other known fast fading transients.
Although there is uncertainty in the explosion scenario due to difficulty in measuring the explosion epoch, we find PS17cke to be a plausible kilonova candidate from the model comparisons.
\end{abstract}


\maketitle


\section{Introduction}
\label{sec:introduction}

Transient phenomena, such as supernovae (SNe), the explosive deaths of massive stars, are renowned for displaying huge diversity in their observed peak luminosity and temporal evolution.
Numerous wide-field survey telescopes are currently in operation, scanning the ever-changing night sky for new and interesting objects that vary in time.
Projects such as the Asteroid Terrestrial-impact Last Alert System \citep[ATLAS,][]{Tonry2018} and the Zwicky Transient Facility \citep[ZTF,][]{Bellm2019} are currently at the forefront of rapid cadence observations, and routinely survey the night sky visible to them every two nights. 
Conversely, the All-Sky Automated survey for Supernovae \citep[ASAS-SN,][]{Shappee2014} uses wider-field, small aperture, telescopes to survey to shallower depths, but with multiple units situated across the globe, it can cover the whole sky each night. 
Deeper surveys that offer higher resolution and multi-colour coverage of smaller sky footprints, such as the Dark Energy Survey \citep[DES,][]{Flaugher2005}, are also competing to discover these types of transients.
Among these programs, the \textit{Panoramic Survey Telescope and Rapid Response System} (Pan-STARRS) combines the depth of a 2\,m telescope with the high survey cadence that a wide-field facility provides and multi-wavelength filter coverage \citep{Chambers2016}.
We will discuss the technical specifics of the Pan-STARRS survey in the next section. 

Since the start of the Pan-STARRS1 Science Consortium surveys in 2009, the facility has produced numerous discoveries and benchmark papers in the time domain era. 
The Pantheon sample of Type Ia SNe is now the leading sample for cosmological constraints \citep{Scolnic2018} and a foundation low redshift sample has been recently published 
\citep{Foley2018}. 
On the more exotic transient side, the Pan-STARRS project has made an impact in the discovery of 
tidal disruption events \citep[TDEs,][]{Gezari2012,Chornock2014,Holoien2019,Nicholl2019} superluminous supernovae \citep[SLSNe,][]{Chomiuk2011,McCrum2015,Lunnan2018}, a population of `fast evolving' transients \citep{Drout2014} and outbursts of stars before core-collapse \citep{Fraser2013}. 

The recently discovered `fast' transients (objects with lightcurves lasting days to weeks), such as AT2018cow \citep{Prentice2018,Perley2019}, AT2018kzr \citep{McBrien2019,Gillanders2020} and SN2019bkc \citep{Chen2019,Prentice2020}, as well as the sample of DES objects discussed by \citet{Pursiainen2018}, present a new parameter space for survey telescopes to probe.
Objects such as these often prove to be intrinsically faint as well, which makes detection more difficult \citep{Srivastav2020}. 
Among the most extreme fast and faint transients that are theoretically predicted, and have subsequently been observed to exist, are kilonovae - the electromagnetic signal emanating from the site of a merger between two neutron stars (NS-NS), or a black hole and neutron star (BH-NS) \citep{Li1998,Rosswog2005,Metzger2010}.
This emission, peaking in the optical and near infrared (NIR) on a timescale of a few days to a week, is caused by the decay of rapid neutron capture \citep[$r$-process,][]{Eichler1989,Freiburghaus1999} elements synthesised during the merger.

The first confirmed kilonova with an unambiguous gravitational wave (GW) signature from a binary neutron star system was discovered in August 2017 \citep{LigoVirgo2017} and ushered in a new era of multi-messenger astrophysics \citep{MMApaper2017}.
In addition to the GW signal, GW170817, an associated short duration gamma ray burst \citep[sGRB,][]{Abbott2017GRB,Goldstein2017} was discovered along with the predicted optical counterpart.
This optical counterpart was given the designation AT2017gfo \citep{Andreoni2017,Arcavi2017,Coulter2017,Chornock2017,Cowperthwaite2017,Drout2017,Evans2017,Kasliwal2017,Lipunov2017,Nicholl2017,Tanvir2017,Pian2017,Troja2017,Smartt2017,SoaresSantos2017,Utsumi2017,Valenti2017}.
In the weeks and months that followed its discovery, observations were undertaken in radio \citep{Hallinan2017}, as well as x-ray \citep{Evans2017,Margutti2017}, all identifying a GRB-like afterglow. 
Since the discovery of this first kilonova, survey telescopes have been attempting to identify more of these events, concurrently and independently of GW triggers. 
These efforts have proved challenging as kilonovae can decline by as many as 2 magnitudes per day in the bluer bands.
The link between binary neutron star mergers and sGRB events had long been suspected, which had prompted detailed follow-up campaigns of sGRBs to search for kilonovae in the fading afterglows \citep{Perley2009,Berger2013,Tanvir2013,Jin2015,Yang2015,Jin2016}.

Now, in the era of gravitational wave observations with LIGO and Virgo \citep{Abbott2016BBH1,Abbott2016BBH2,Abbott2017BBH3}, we are able to conduct follow-up of the localised sky areas released in the wake of publicly announced GW signals.
With Pan-STARRS, we conducted follow-up of the binary black hole mergers detected during the first and second LIGO observing runs \citep[O1 and O2,][]{Smartt2016BBH1,Smartt2016BBH2}.  
In the first 12 to 36 hours of follow-up of AT2017gfo, our Pan-STARRS observations, combined with image subtraction, provided the quantitative early evidence for rapid photometric fading, mainly due to the pre-existing templates \citep{Chambers2016}. 
The key to this was having the Pan-STARRS1 Science Consortium 3$\pi$ reference images, which provide a template for image subtraction at declinations above $\delta = -30^{\circ}$.
For the joint LIGO-Virgo observing run, O3, the Hanford and Livingston units have become sensitive to detecting gravitational waves from binary neutron star mergers out to angle averaged distances of $100-120$\Mpc, where Virgo is sensitive to approximately 50\Mpc. The horizon distance for the two LIGO detectors is closer to 160\Mpc\ (distance modulus $\mu=36$), and we would anticipate a kilonova peaking at an absolute magnitude of $-14 > M > -16$ to be detected at apparent magnitudes in the range $20 > m > 22$. This is comfortably above the typical Pan-STARRS detection limits in its filter system (discussed in the next section) with exposure times of $30-960$ seconds. 

The discovery of AT2017gfo, associated with GW170817, indicates that kilonovae with peak absolute magnitudes of $M\simeq-15.5$ exist in the local Universe and, irrespective of GW detections, should be discoverable by ZTF, ATLAS and Pan-STARRS.
Its discovery also changed how various groups conducted follow-up of faint and fast transients as objects of this magnitude and decline rate until now were poorly characterised and few had been observed.
It also posed the question of which instruments would best detect these rapidly evolving and lower luminosity events \citep{Scolnic2018}.
Such luminosities are similar to faint SNe \citep[such as faint Type II-P and the extreme end of the SN Iax population,][]{Pastorello2004,Spiro2014,Srivastav2020} and hence the two factors which are preventing independent discovery at optical to NIR wavelengths are the intrinsic rates and the cadence strategies of the surveys. 
The known sGRB population has a median redshift of $z\simeq0.48$ \citep{Berger2014} with the lowest redshift sGRB apart from that associated with GW170817 being $z\simeq0.111$ \citep{Berger2014}.
Hence, finding kilonovae in the fast fading afterglows of sGRBs is challenging, given that their distance implies they would typically be fainter than $20^{\mathrm{th}}$ magnitude.

Previous searches for fast transients were conducted in the Pan-STARRS1 Science Consortium Medium Deep Survey fields by \citet{Drout2014} and \citet{Berger2013ps1}.
\citet{Drout2014} discovered rapidly evolving objects in the range $-16.5 > M > -20.0$, which  declined by at least 1.5 magnitudes in less than 25 days. 
These are several magnitudes brighter than AT2017gfo and declined much more slowly.
In its current main survey mode, the Pan-STARRS twin telescope system carries out a survey for near earth objects (NEOs), funded by NASA.
In a companion paper to this (Smartt et al. 2020, in prep.), we introduce the Pan-STARRS Search for Kilonovae (PSSK) program being undertaken to identify kilonovae and other fast and faint transients with the Pan-STARRS NEO data.
These images typically reach ${\sim}21.5 - 22$ for each 45\,s exposure and cover around 1000\,square degrees per night.
We now announce all candidates publicly \citep{Smartt2019AN} and have performed an archival search through all previously taken Pan-STARRS data prior to August 2019, which we discuss in this paper. 

Smartt et al. (2020, in prep.) presents the analysis of 1975 transients within 200\Mpc\ and finds two rapidly declining transients that were worthy of consideration as kilonovae candidates. 
In this paper we analyse these two objects - PS15cey and PS17cke - in detail.
In Section \ref{sec:pssk}, we outline the technical observing specifications of Pan-STARRS1, the primary telescope operating under the Pan-STARRS survey, and outline briefly the work being done as part of PSSK.
In Sections \ref{sec:ps15cey} and \ref{sec:ps17cke}, we discuss the prospective kilonova candidates identified through the archival search of PSSK.
In Section \ref{sec:discussion}, we discuss the implications and plausible progenitor scenarios for these events before concluding in Section \ref{sec:conclusions}. Throughout this paper, we adopt cosmology of $\mathrm{H}_0 = 70$\kmsMpc, with $\Omega_m = 0.3$ and $\Omega_\Lambda = 0.7$.

\section{Pan-STARRS and the search for kilonovae}
\label{sec:pssk}

The Pan-STARRS1 system \citep{Chambers2016} comprises a 1.8\,m telescope located at the summit of Haleakala on Maui, Hawaii, with a 1.4 gigapixel camera, called GPC1, mounted at the Cassegrain $f$/4.4 focus.
GPC1 is composed of sixty orthogonal transfer array devices, each of which has a detector area of $4,846 \times 4,868$ pixels. These pixels measure 10\,$\mathrm{\mu}$m in size giving a focal plane of 418.88\,mm in diameter, or equally 3.3$^{\circ}$.
This corresponds to field-of-view of 7.06 square degrees.
The Pan-STARRS1 filter system (\grizy) is similar to that of SDSS \citep{Abazajian2009}, but it also includes a composite \gps\rps\ips\ filter called `wide' or \wps, and is described in detail by \citet{Tonry2012} and \citet{Chambers2016}.
In nightly survey operations PS1 typically observes in quads of 45\,s \wps\ exposures, reaching \wps$\lesssim$22 magnitudes.

Images taken from Pan-STARRS1 are processed immediately with the Image Processing Pipeline \citep[IPP,][]{Magnier2016data}.
Through the Pan-STARRS1 3$\pi$ Steradian Survey data \citep{Chambers2016}, we have a ready made template of the whole sky north of $\delta = -30^{\circ}$, and proprietary \ips\ data in a band between $-40^{\circ} < \delta < -30^{\circ}$, giving a reference sky in the \ips\ band down to this lower declination limit.
Frames are astrometrically and photometrically calibrated with the standard Image Processing Pipeline steps \citep{Magnier2016data,Magnier2016pixel,Magnier2016phot}.
The Pan-STARRS1 3$\pi$ reference sky images are subtracted from these frames \citep{Waters2016} and photometry is carried out on the resulting difference images \citep{Magnier2016pixel}.
The detections are then processed through the Pan-STARRS Transient Science Server (PTSS) hosted on a dedicated computing cluster at Queen's University Belfast.
The processing involves assimilating the detections into objects and lightcurves, removing false positives, identifying asteroids, variable stars and AGN, and cross-matching with known galaxy catalogues to determine the most likely host and assign a photometric or spectroscopic redshift, if known. 
The Pan-STARRS Transient Science Server is also described in our companion paper (Smartt et al. 2020, in prep.). 

Smartt et al. 2020 (in prep.) also introduces the Pan-STARRS search for kilonovae (PSSK).
From a transient database of over $20,000$ detection, 1975 were found to be plausibly associated with galaxies within 200\Mpc. 
We were able to rule out 1758 as being either confirmed or likely supernovae from a combination of spectral classifications, absolute magnitudes and lightcurve evolution.
There were 215 (11\%) that had only one data point and could not be classified further. 
This left 2 which have unusual, fast declining lightcurves - PS15cey, a rapidly declining object with detections on two epochs and in two bands, and PS17cke, an intrinsically faint object only detected on one epoch but with a recent post-discovery non-detection that constrained the decline rate to be similar to AT2017gfo. 

We also use ATLAS to search for transients within a fixed local volume \citep{Smith2020}, to shallower depths than Pan-STARRS but with significantly higher cadence.
ATLAS is a twin 0.5\,m telescope system with units based on Haleakala and Mauna Loa, Hawaii 
\citep{Tonry2018}.
The Wright-Schmidt telescopes each have a 10\,k\,$\times$\,10\,k detector, providing a 28 square degree field of view.
ATLAS robotically surveys the sky in cyan (\textit{c}) and orange (\textit{o}) filters that are broadly similar to composite Pan-STARRS/Sloan Digital Sky Survey (SDSS) \textit{gr} and \textit{ri} filters, respectively.
The system covers the whole sky visible from Hawaii every two nights with a pattern of 4$\times$30\,s dithered exposures per night, spaced approximately equally within a 1 hour period.
The survey images are processed as described by \cite{Tonry2018} and are photometrically and astrometrically calibrated immediately. 
Each image has a deep reference sky image subtracted and transient sources detected as described in 
\citep{Smith2020}.
We make use of ATLAS data to supplement these two Pan-STARRS discoveries. 

\section{Data analysis for PS15cey}
\label{sec:ps15cey}

\subsection{Discovery and initial photometry}

\begin{figure*}
    \centering
    \includegraphics[width=\linewidth]{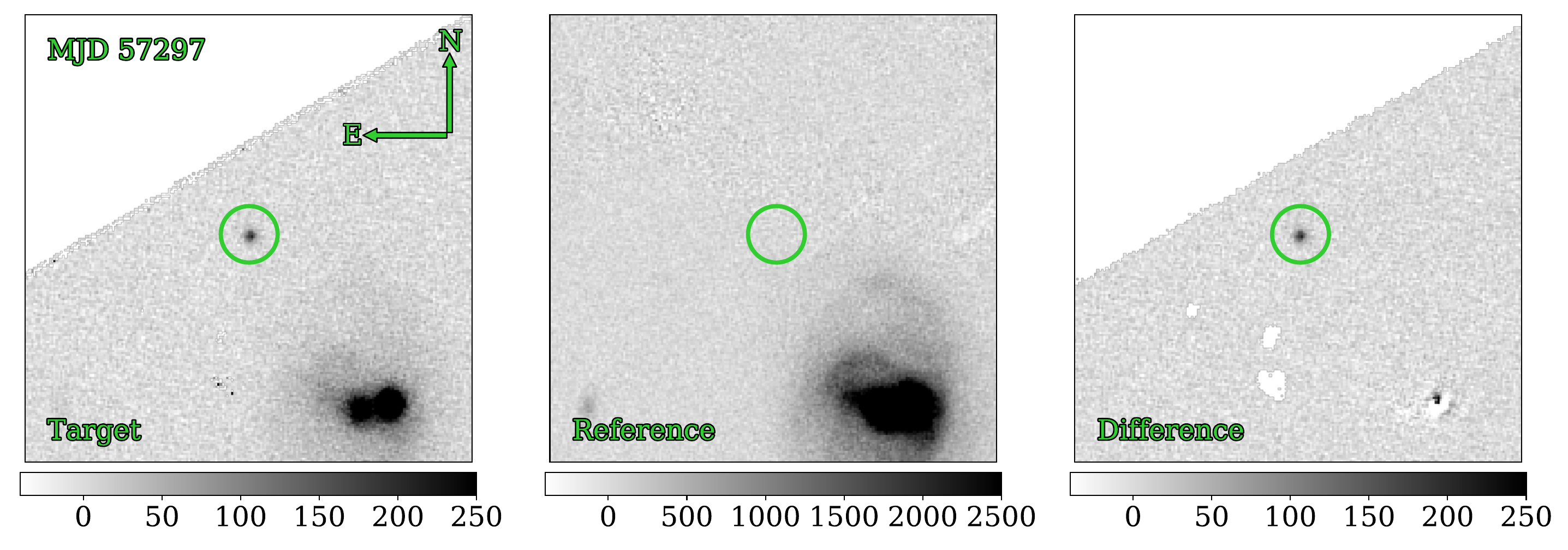}
    \caption{The Pan-STARRS1 discovery image triplet of PS15cey taken on the night of MJD 57297 in \rps. \textbf{Left}: The target frame acquired on MJD 57297.350 (2015-10-02 08:23:42 UTC) in \rps. \textbf{Centre}: The reference wallpaper image used for subtraction. \textbf{Right}: The resulting subtraction showing the clear detection of PS15cey.
    } 
    \label{fig:ps15cey_triplet}
\end{figure*}

PS15cey (Figure \ref{fig:ps15cey_triplet}) was discovered on MJD 57297.350 (2015-10-02 08:23:42 UTC) 
with a magnitude of \rps=$19.17 \pm 0.03$ at location RA = 22:37:00.81 and Dec = -16:13:47.5.
PS15cey was observed a total of eight times by PS1 - four times in \rps\ on the night of MJD 57297 
and four times in \ips\ on the subsequent night, MJD 57298, which average to \ips$= 19.64 \pm 0.03$.
The last previous non-detection before discovery occurred on MJD 57246.505 (2015-08-12 12:07:12 UTC) at \wps\ > 22.2. 
The closest non-detection in time following the event is a shallow image from MJD 57353.200 (2015-11-27 04:48:00 UTC) at \zps\ > 18.9, with the next PS1 observation of the field being over 200 days later.
No other detections of this transient were seen in PS1 data over the period MJD 56459 to MJD 58808.
The foreground extinction in this direction is low, with $A_r=0.1$ and $A_i=0.08$\,mag \citep{Schlafly2011}, implying an intrinsic colour difference of \rps-\ips$=-0.50\pm0.05$\,mag. 
This could either be due to a rapid decline in luminosity of the transient, a colour difference or a combination of both.
We show our photometry of PS15cey in the top panel of Figure \ref{fig:plotPS15ceyLightcurve}.

\subsection{Spectra}

\begin{figure}
    \centering
    \includegraphics[width=\linewidth]{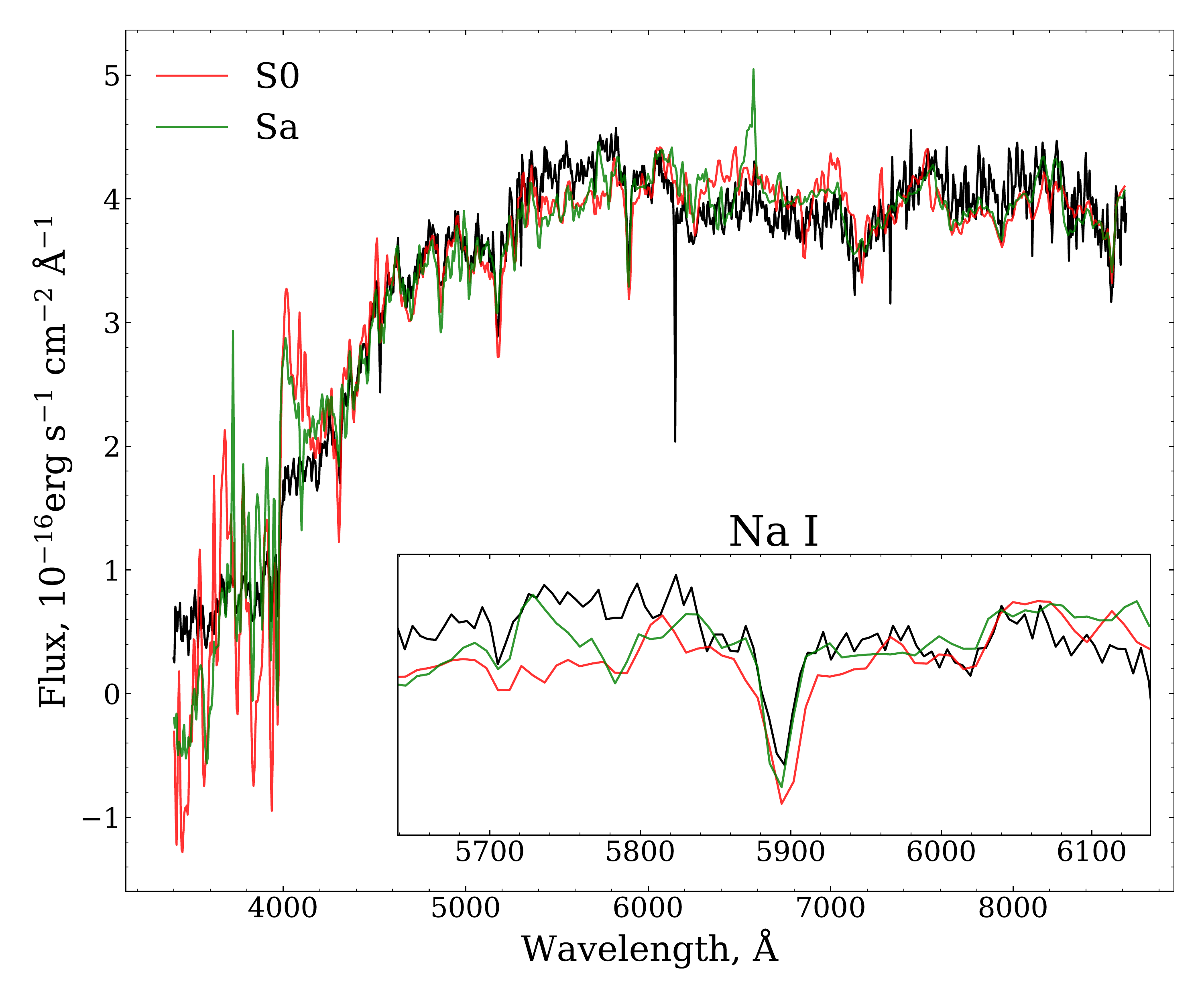}
    \caption{The spectrum of the host of PS15cey taken MJD 58816.066 (2019-11-29 01:35:41 UTC) with NTT:EFOSC2. For comparison, we show scaled spectra of typical S0 and Sa galaxies in the same wavelength range \citep{Kinney1996}. From this spectrum we have inferred a redshift to the host of $z = 0.0717 \pm 0.0006$, with the inset plot showing the \NaI\ line which was used in the measurement of this redshift.
    }
    \label{fig:plotHostWithTemplates}
\end{figure}

Cross-matching with galaxy catalogues suggests a potential host in the galaxy WISEA J223659.87-161406.7, though the host is identifiable in several catalogues. Within LEDA, this galaxy is recognised as PGC900758, and, according to the GLADE catalogue, has a redshift of $z=0.038$. This is likely a photometric redshift however, as a spectroscopic source cannot be corroborated, and hence is not reliable enough to draw conclusions from. 
Owing to this uncertainty on the redshift, a spectrum of the host galaxy was taken with EFOSC2 at the New Technology Telescope (NTT), using Grism \#13 (wavelength coverage $3685-9315$\AA), on MJD 58816.066 (2019-11-29 01:35:41 UTC) as part of the advanced Public ESO Spectroscopic Survey for Transient Objects (ePESSTO+).
The data were reduced with the PESSTO pipeline \citep{Smartt2015} and the extended galaxy spectrum was extracted. 
The spectrum has identifiable emission and absorption features, including the \NaI\ lines and \CaII\ H\&K lines, as well as some $H\alpha$ emission.
From these lines, we can measure a redshift to the host of $z=0.0717 \pm 0.0006$.
This corresponds to a luminosity distance of 320\Mpc\ for our chosen cosmology.
We estimate that PS15cey is offset by 32\farcs4 from its host, or equivalently a physical separation of approximately 50.3\kpc.
At this redshift, the host spectrum resembles those of typical S0 or Sa type galaxies \citep{Kinney1996}, as can be seen in Figure \ref{fig:plotHostWithTemplates}.
Given this redshift, as well, the absolute magnitude at peak of the existing photometry would be $-18.40$ in \rps\ and $-17.89$ in \ips. 

\subsection{Decline}

If we assume that the transient did not fade significantly between the two epochs, then \rps-\ips$=-0.50\pm0.05$ implies a very blue colour for the spectral energy distribution (SED).
It would require a blackbody with an effective temperature of $26,500 \pm 3500$\,K to provide this blue colour.
This is extremely hot, and such temperatures are usually only seen at shock breakout, with higher temperatures reached during the post-shock cooling phase \citep{Yaron2017}.
However, if these data were the signature of shock breakout, then a rising supernova should have been visible in the days and weeks that followed.
There is no detection on MJD 57353.200 (2015-11-27 04:48:00 UTC), or +51 days (from the rest frame) after discovery in Pan-STARRS, with a limit of \zps$>18.9$. 
In ATLAS, the closest observation is on MJD 57378, and forced photometry\footnote{Forced photometry is a technique used to measure the flux at the position of transient in a difference image where it does not have a clear, $5\sigma$ significance detection. The technique uses the spatial position of the transient in a high significance detection image and the point spread function (PSF) of the image from brighter stars. A PSF model is built exactly at the known position of the transient and is fit to the flux in the difference image. This allows flux to be reliably recovered below the $5\sigma$ limit, while ensuring that the flux is measured at the known position of the transient with a suitable PSF.} on the ATLAS survey data shows no transient to typical magnitudes of $o > 18.3$.
We quote the flux in Table \ref{tab:my_label} as a flux measurement, unlike logarithmic magnitude, is meaningful even if small or negative.

Hence, it is likely that the two photometric measurements are a signature of a real and rapid decline. 
The most recent non-detections occurred on MJD 57246.505 (2015-08-12 12:07:12 UTC) at ${>}22.2$ in \wps\ prior to discovery.
A decline rate of $-0.5$\,mag\,day$^{-1}$ is comparable to both the NS-NS merger AT2017gfo and the possible WD-NS merger AT2018kzr \citep{McBrien2019,Gillanders2020}.
However, the secure redshift of the host galaxy PGC900758 places its peak absolute magnitude at least two magnitudes brighter than AT2017gfo, which is more similar to the peak brightness of AT2018kzr.
We note that at the time of discovery by PS1, the intermediate Palomar Transient Factory \citep{Rau2009,Law2009,Kulkarni2013ATel} was observing also, but has no observations of the field of PS15cey within the months surrounding its discovery\footnote{Private communications with Russ Laher, Frank Masci and Adam Miller}.

\begin{figure}
    \centering
    \includegraphics[width=\linewidth]{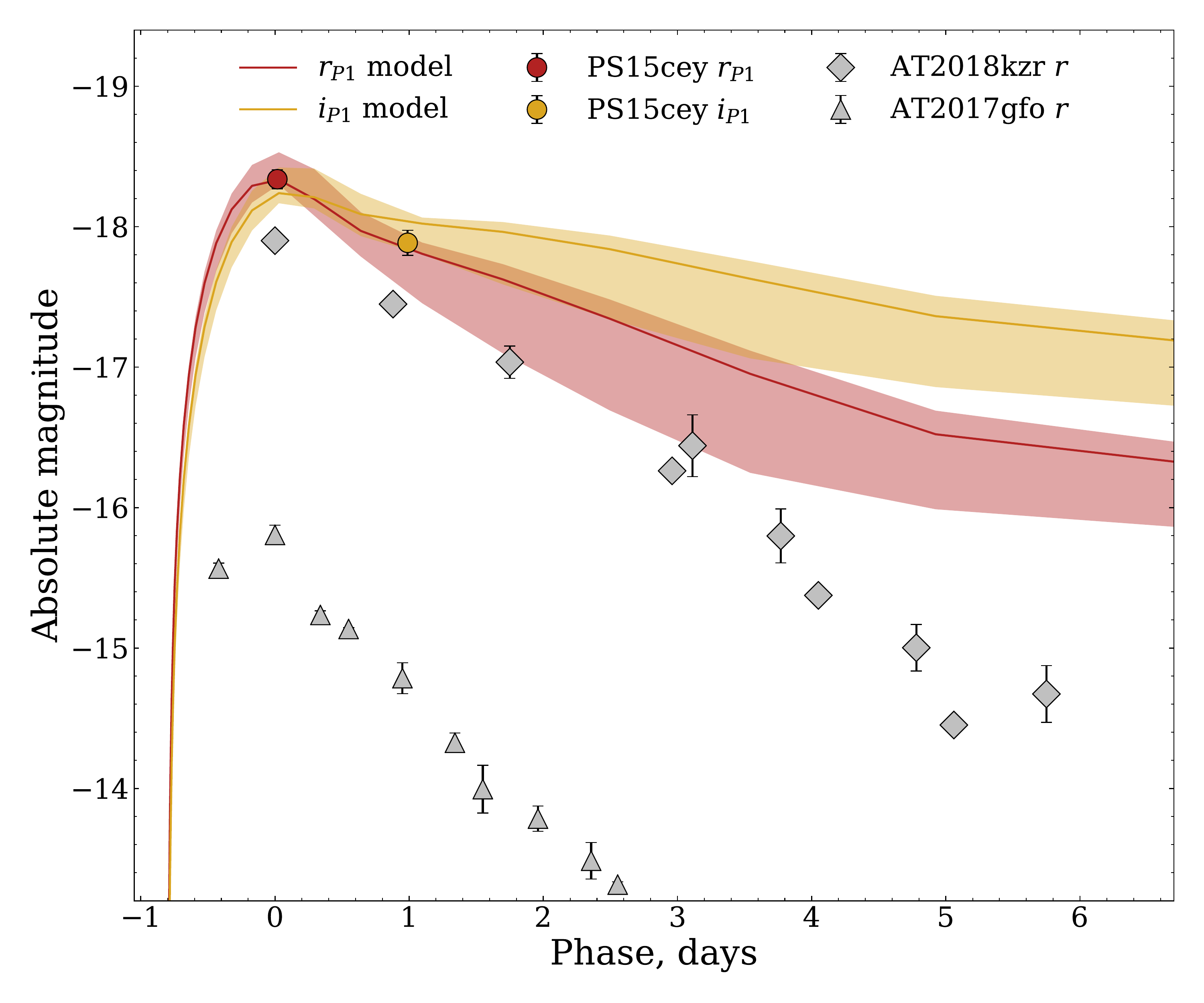}
    \caption{The Pan-STARRS1 lightcurve of PS15cey with detections averaged in \rps\ (red) and \ips\ (yellow) across each night of observing. We overlay on this the absolute lightcurves of AT2018kzr \citep[diamonds,][]{McBrien2019,Gillanders2020} and AT2017gfo \citep[triangles,][]{Andreoni2017,Arcavi2017,Chornock2017,Cowperthwaite2017,Drout2017,Evans2017,Kasliwal2017,Tanvir2017,Pian2017,Troja2017,Smartt2017,SoaresSantos2017,Utsumi2017,Valenti2017}, in addition to BH-NS kilonova models, fit each data point. We discuss the model formulation and interpretation in Section \ref{sec:ps15cey:disc}. Phases are given with respect to the discovery of PS15cey.}
    \label{fig:plotPS15ceyLightcurve}
\end{figure}

\begin{table*}
    \centering
     \caption{Photometry and flux measurements of PS17cke. The flux is in $\mathrm{\mu}$Jy and is a forced photometry measurement at each epoch in the difference frames. Where a magnitude limit is quoted, it is the 5$\sigma$ limit based on $\sigma$ being the uncertainty in the forced photometry flux. The flux errors are the standard deviation of multiple frames where they exist (ATLAS and PS1), and the exposure times are total sum of frames on that night.
    }
    \begin{tabular}{llllccc}\hline
    MJD          & Epoch (days) & Facility & Filter & Exposure time (s) & AB magnitude     & Flux ($\mathrm{\mu}$Jy)  \\\hline 
    57828.569    & -30.739      & ATLAS    & $o$    &  9 $\times$ 30    & ${>}17.19$       & -88   $\pm$ 96            \\
    57830.566    & -28.742      & ATLAS    & $o$    &  4 $\times$ 30    & ${>}17.86$       & -76   $\pm$ 52            \\
    57832.551    & -26.757      & ATLAS    & $o$    & 14 $\times$ 30    & ${>}17.29$       & -79   $\pm$ 88            \\
    57833.551    & -25.757      & ATLAS    & $o$    &  4 $\times$ 30    & ${>}14.36$       & -628  $\pm$ 1306          \\
    57834.549    & -24.759      & ATLAS    & $o$    &  4 $\times$ 30    & ${>}17.47$       & -67   $\pm$ 74            \\
    57836.553    & -22.755      & ATLAS    & $c$    &  5 $\times$ 30    & ${>}19.91$       & 9     $\pm$ 8             \\
    57837.529    & -21.779      & ATLAS    & $o$    &  8 $\times$ 30    & ${>}15.63$       & -448  $\pm$ 405           \\
    57838.534    & -20.774      & ATLAS    & $o$    &  2 $\times$ 30    & ${>}18.70$       & -72   $\pm$ 24            \\
    57840.511    & -18.797      & ATLAS    & $c$    &  2 $\times$ 30    & ${>}19.54$       & -36   $\pm$ 11            \\
    57841.416    & -17.892      & PS1      & \wps\  &  180              & ${>}21.40$       & 6     $\pm$ 2             \\
    57842.524    & -16.784      & ATLAS    & $o$    &  4 $\times$ 30    & ${>}17.36$       & -344  $\pm$ 82            \\
    57844.516    & -14.792      & ATLAS    & $c$    &  4 $\times$ 30    & ${>}18.49$       & -5    $\pm$ 29            \\
    57845.506    & -13.802      & ATLAS    & $o$    &  2 $\times$ 30    & ${>}17.53$       & -214  $\pm$ 70            \\
    57846.503    & -12.805      & ATLAS    & $o$    &  4 $\times$ 30    & ${>}17.09$       & -501  $\pm$ 105           \\
    57846.506    & -12.802      & PS1      & \wps\  &  360              & 21.09 $\pm$ 0.16 & 13    $\pm$ 2             \\
    57849.480    & -9.828       & ATLAS    & $o$    &  4 $\times$ 30    & ${>}17.95$       & -120  $\pm$ 48            \\
    57857.481    & -1.827       & ATLAS    & $o$    &  3 $\times$ 30    & ${>}15.79$       & -182  $\pm$ 348           \\
    57858.473    & -0.835       & ATLAS    & $o$    &  4 $\times$ 30    & ${>}17.73$       & -104  $\pm$ 58            \\
    57859.308    & +0.000       & PS1      & \wps\  &  180              & 19.69 $\pm$ 0.02 & 47    $\pm$ 1             \\
    57861.537    & +2.229       & ATLAS    & $o$    &  4 $\times$ 30    & ${>}18.11$       & -1898 $\pm$ 41            \\
    57862.288    & +2.980       & UH2.2m   & $V$    &  20               & $-$              & $-$                       \\
    57862.542    & +3.234       & ATLAS    & $o$    &  4 $\times$ 30    & ${>}17.97$       & -215  $\pm$ 47            \\
    57862.857    & +3.549       & SALT     & \wps\  &  5                & $-$              & $-$                       \\
    57863.369    & +4.061       & PS1      & \wps\  &  90               & ${>}20.96$       & 4     $\pm$ 3             \\
    57864.765    & +5.457       & SALT     & \wps\  &  5                & $-$              & $-$                       \\
    57865.258    & +5.950       & NTT      & $V$    &  200              & 21.86 $\pm$ 0.18 & 5     $\pm$ 2             \\
    57865.521    & +6.213       & ATLAS    & $o$    &  4 $\times$ 30    & ${>}18.00$       & -53   $\pm$ 46            \\
    57866.520    & +7.212       & ATLAS    & $o$    &  4 $\times$ 30    & ${>}14.29$       & 282   $\pm$ 1395          \\
    57868.505    & +9.197       & ATLAS    & $c$    &  2 $\times$ 30    & ${>}18.57$       & -25   $\pm$ 27            \\
    57869.511    & +10.203      & ATLAS    & $o$    &  4 $\times$ 30    & ${>}17.25$       & -170  $\pm$ 91            \\
    57870.331    & +11.023      & PS1      & \wps\  &  180              & ${>}20.96$       & -1    $\pm$ 3             \\
    57876.400    & +17.092      & ATLAS    & $o$    &  1 $\times$ 30    & ${>}14.62$       & -312  $\pm$ 1026          \\
    57881.374    & +22.066      & ATLAS    & $o$    &  4 $\times$ 30    & ${>}18.11$       & -33   $\pm$ 41            \\
    57882.393    & +23.085      & ATLAS    & $o$    &  6 $\times$ 30    & ${>}17.19$       & -91   $\pm$ 96            \\
    57883.477    & +24.169      & ATLAS    & $o$    &  4 $\times$ 30    & ${>}16.42$       & -3    $\pm$ 196           \\
    57884.398    & +25.090      & ATLAS    & $o$    &  9 $\times$ 30    & ${>}17.38$       & -52   $\pm$ 81            \\
    57886.375    & +27.067      & ATLAS    & $o$    &  4 $\times$ 30    & ${>}17.36$       & -116  $\pm$ 82            \\
    57892.419    & +33.111      & ATLAS    & $o$    &  5 $\times$ 30    & ${>}19.17$       & -32   $\pm$ 16            \\\hline 
    \end{tabular}
    \label{tab:my_label}
\end{table*}

\section{Data analysis for PS17cke}
\label{sec:ps17cke}

\subsection{Discovery and initial photometry}

\begin{figure*}
    \centering
    \includegraphics[width=\linewidth]{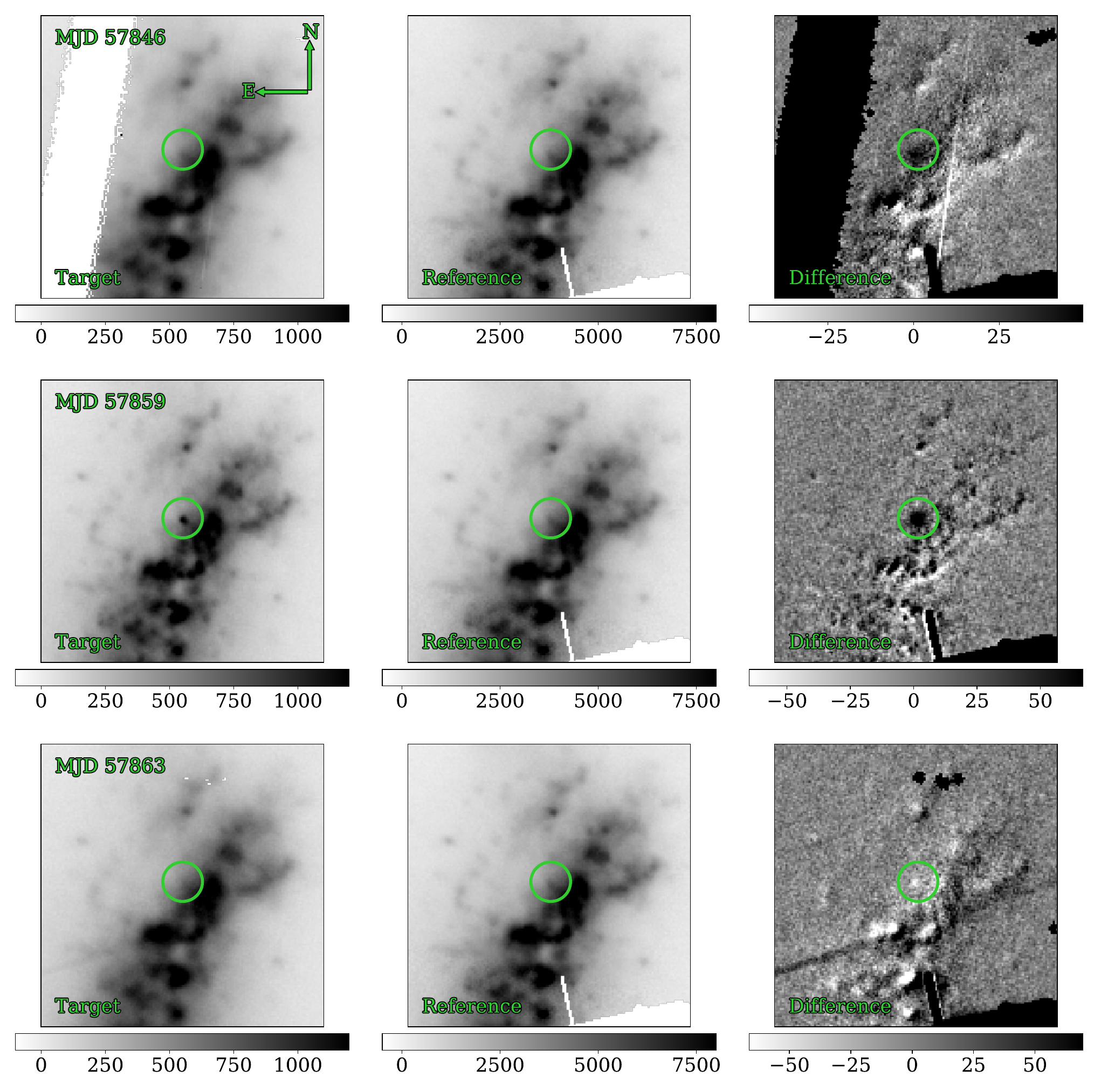}
    \caption{The Pan-STARRS1 discovery image triplets of PS17cke at the three epochs of our photometry. From left to right the panels show the target frame for a particular epoch, then the reference frame used for subtraction (which is the same for all epochs), then finally the resultant subtraction frame. \textbf{Top}: The most recent pre-discovery detection and subtraction frames of PS17cke made on MJD 57846.506 (2017-04-03 12:08:41 UTC) via a forced photometry measurement in \wps. \textbf{Middle}: The discovery epoch detection and subtraction frames made on MJD 57859.308 (2017-04-16 07:24:06 UTC) in \wps. \textbf{Bottom}: The most recent post-discovery non-detection and subtraction frames, included to emphasises the rapid decline of PS17cke, made on MJD 57863.369 (2017-04-20 08:51:35 UTC) in \wps.
    }
    \label{fig:ps17cke_triplet}
\end{figure*}

PS17cke (Figure \ref{fig:ps17cke_triplet}) was discovered at \wps=$19.69 \pm 0.04$ on MJD 57859.308 (2017-04-16 07:24:06 UTC) at location RA = 12:34:18.90 and Dec = +06:28:27.0.
The object was only detected on one night in all four exposures taken in \wps.
It was registered on the Transient Name Server\footnote{\url{https://wis-tns.weizmann.ac.il/}} and given the IAU designation AT2017des \citep{Smith2017ATel}.
We note that PS17cke was discovered approximately 4 months before AT2017gfo, so would have been overlooked as a kilonova candidate at the time.
PS17cke lies in the nearby star-forming host galaxy NGC 4532.
It is in a high surface brightness region of the irregular host, making image subtraction difficult.
NGC 4532 has a heliocentric velocity recorded in the NASA/IPAC Extragalactic Databse (NED) of 2061\kms\ or $z = 0.00671 \pm 0.00001$ \cite[e.g.][]{Binggeli1985}, which would correspond to a distance of 27\,Mpc for our adopted cosmology.
However, it lies in the direction of the Virgo galaxy cluster and is either in the cluster or its recessional velocity is significantly perturbed by Virgo's mass.
The correction for Virgo, Great Attractor and Shapley used in NED \cite[by][]{Mould2000}, implies a kinematic distance of 14\,Mpc while direct estimates (Tully-Fisher and H\,I correlation methods) report distances between 10 - 20\,Mpc \citep[see][for two of the extremes]{Tully1984,Yasuda1997}.
NGC 4532 is toward the subcluster B around M49 in the southern regions of Virgo, to which \citet{Mei2007} estimate a distance of 16\,Mpc. 
We hence adopt $15 \pm 5$\,Mpc, an equivalent distance modulus of $\mu = 30.9 \pm 0.8$ as the distance to NGC 4532, and will propagate the distance uncertainty through the discussions of the nature of the transient.

In the weeks surrounding the discovery of PS17cke, the field was observed a total of 36 times across 7 epochs by PS1.
Typical 5$\sigma$ limiting magnitudes of the individual 45 second images in this period were between 
$21<$\wps$<22$, but the high surface brightness of NGC 4532 reduces the sensitivity of the difference images at the location of the transient. 
None of these difference images, nor any historical image from Pan-STARRS1, produced an independent 5$\sigma$ positive detection at the location of PS17cke.
The lack of detection just 4.06 days later was the reason this object was uncovered from the sample of 1975 objects within 200\,Mpc in Smartt et al. (2020 in prep.). 
To further constrain detections and detection limits, we forced photometry at the position of all of the Pan-STARRS1 45\,s difference images at the location of PS17cke from MJD 57805.481 (2017-02-21 11:32:38 UTC) to MJD 58879.594 (2020-01-31 14:15:21 UTC).
A point-spread-function (PSF) was calculated for each target image and a flux measurement using this PSF was forced at the astrometric position of PS17cke.
All of the images were in the \wps\ filter and the data typically consist of a sequence $4 \times 45$\,s exposures per night, but on some nights a sequence of $8 \times 45$\,s was available.
All images were taken in normal survey mode and were not targeted observations of PS17cke itself.
The forced photometry is reported in microJansky units in Table \ref{tab:my_label} on a nightly basis.
The nightly forced flux is the average of the flux measurements from the individual images, with the error being the standard deviation.
These were converted to magnitudes if the significance of detection was greater than 5$\sigma$. 

In addition we also co-added the individual 45\,s frames to visually inspect them for any sign of significant positive flux.
The transient is not significantly detected in the images $-17.9$ days before discovery and +4.1 days after to the limits quoted in Table \ref{tab:my_label}.
However, positive flux appears in the frames from MJD 57846.485 (2017-04-03 11:38:24 UTC, 12.8 days before discovery).
On this night a $8\times45$\,s sequence of frames was taken.
Each image had a positive flux residual of less than 5$\sigma$ significance but the average and standard deviation of the eight flux measurements implies a combined ${\sim}6.5\sigma$ detection corresponding to magnitude \wps=$21.09 \pm 0.16$.
Inspecting the forced photometry lightcurve and image data across all epochs, indicate that the high surface background leaves positive and negative residuals in the difference images at similar levels to the `source' detected in this -12.8 day epoch.
Hence, it is not completely secure if this is a real detection of transient flux.
To check the reality of this forced positive flux, we co-added the 8 frames and applied aperture photometry, but allowed the centroid to vary.
This produces a measurement of positive flux of similar significance to that in the combined difference image measurements at a position which is 0\farcs23 offset from the centroid of the high signal-to-noise detection on the discovery epoch, indicating that this was not ambiguously transient flux. 
We carried out two further tests to determine if this flux was real.
A higher quality reference stack was constructed with all PS1 \wps\ images outside the timeframe of possible detection, giving a deep 660\,s frame made from $4\times30$\,s and $12\times45$\,s images.
We applied the standard PS1 IPP difference imaging pipeline to this, combined with forced photometry.
The results are plotted in Figure\,\ref{fig:atlas_lc_ps17cke_snr5}.
Again, we measured similar positive residuals in the difference images at the position of PS17cke on the $-12.8$ day epoch.
However in all cases there is a consistent residual and positive flux measured at this position with an average of around 6$\mathrm{\mu}$Jy.
We attribute this to imperfect image subtraction due to a combination of high surface brightness, probable colour gradients within the PSF and the wide \wps\ filter employed.
Inspection of the image frames after subtraction indicated this residual flux was likely to be residual diffuse flux rather than point like.
We also used exactly the same data (i.e. the deep PS1 reference stacks) to run the \texttt{photpipe} difference imaging pipeline \citep{Rest2005}.
This has been employed for accurate photometric measurements in PS1 data by \citet{Rest2014} and \citet{Scolnic2018}, with excellent results.We find similar results to those from the IPP difference imaging.
There are some positive residuals but they are significantly variable within a night and across the 45\,s frames.
A higher positive flux is again found on the $-12.8$ day epoch, but we do not find it to be a convincing PSF-like source.
Given that we are detecting an average background residual of 6$\mathrm{\mu}$Jy above zero, the excess flux in the combined $-12.8$ day epoch is around 3$\sigma$ above this.
For completeness, we have attempted the subtraction with \texttt{hotpants} \citep{Becker2015}.
The resulting difference images it produces contain some flux at this early epoch of a similar significance to the \texttt{photpipe} subtractions.
In the discussion of the nature of the object, we consider the two possibilities of this being a real detection and using it as an upper limit.

\subsection{Spectra}

Four attempts at taking a spectrum of PS17cke were made, but in all cases there is no visible source in the spectroscopic 2D images that could be identified and extracted.
The first attempt was with the SuperNova Integral Field Spectrograph \citep[SNIFS,][]{Lantz2004} on the University of Hawaii's 2.2\,m telescope on MJD 57862.288 (2017-04-19 06:54:41 UTC).
The object was not identifiable in a 20\,s acquisition image, so the observation was aborted and no spectrum was taken. 
Soon after, an observation was made by the Southern African Large Telescope (SALT) with the Robert Stobie Spectrograph (RSS) on MJD 57862.857 (2017-04-19 20:34:07 UTC).
Although identification of the object was not clear in a 5\,s acquisition image, the spectroscopic observation proceeded.
The SALT:RSS spectrum totalled an integration time of 2000\,s.
There are clearly two bright objects or regions in the spectrometer slit (with position angle 113.0\degree\ east from north), which we identify as the bright star USNO-B1 0964-0213613 and a bright knot in the galaxy NGC 4532 some 45\farcs6 away.
These were separated by exactly 90.4 pixels on both the acquisition image and the 2D spectrum and hence we are confident of the slit position. 
This means the 1\farcs5 slit employed in the observations missed the position of PS17cke. 
A subsequent attempt was made with SALT:RSS to observe PS17cke on MJD 57864.765 (2017-04-21 20:45:24 UTC) with the total integration time of 2300\,s.
Similarly, the same two objects could be identified on the 2D spectroscopic frame and the acquisition image and the transient was unlikely to have been included in the slit. 

A further spectrum was attempted on MJD 57865.258 (2017-04-22 06:11:56 UTC) with EFOSC2 at the New Technology Telescope within the ePESSTO program \citep{Smartt2015}. 
Three acquisition images were obtained to identify the target, two with an exposure time of 40\,s and one with 120\,s, but the target is not immediately obvious nor resolvable in any of these.
The acquisition images were taken in the $V$ band and have a measured zero-point of 25.33, based on calibrations to reference stars in the Pan-STARRS catalogue, colour corrected to the Johnson-Cousins filter system.
An attempt to observe the spectrum was made, with a 1000\,s exposure with Grism \#13 (wavelength coverage $3685-9315$\AA).
We do not believe that this spectrum had properly acquired the target due to the faintness of the object amongst the density of resolvable point sources in the host galaxy.
The spectrum extracted shows a blue continuum with narrow emission lines, most likely from the host, including H$\alpha$, [O III] and [S II].
We have measured a secure host redshift of $0.00682 \pm 0.00053$ from the centroids of these lines, by fitting Gaussian profiles to each.

\subsection{Later and forced photometry}

\begin{figure*}
    \centering
    \includegraphics[width=\linewidth]{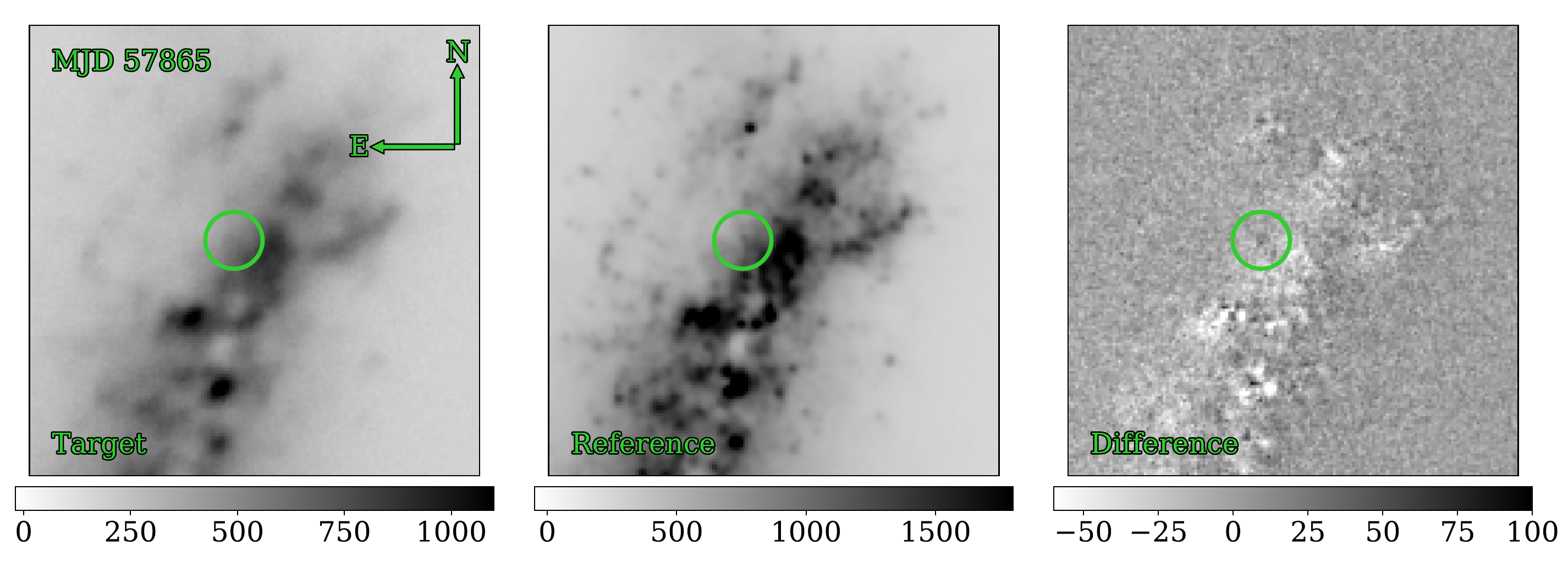}
    \caption{An NTT:EFOSC2 image triplet of PS17cke. \textbf{Left}: The target frame composed of a co-add of the acquisition images taken for the MJD 57865.258 (2017-04-22 06:11:56 UTC) spectrum in Johnson-$V$. \textbf{Centre}: The reference frame used for image subtraction, acquired on MJD 58872.319 (2020-01-24 07:39:42 UTC). \textbf{Right}: The resulting subtraction. Detection of PS17cke is difficult by manual inspection, but a photometric magnitude can be measured from this frame.
    }
    \label{fig:makeSubtractedFrameImage}
\end{figure*}

In order to determine if PS17cke has flux present in the NTT:EFOSC2 200\,s acquisition image, we took a deep reference frame to use for image subtraction.
We observed a set of $12\times100$\,s images with the NTT:EFOSC2 on MJD 58872.319 (2020-01-24 07:39:42 UTC) and produced an aligned, co-added image as shown in Figure \ref{fig:makeSubtractedFrameImage}. 
This was subtracted from the MJD 57865 acquisition exposures to identify any net flux from the object obscured by the host.
There is evidence of some positive flux in the difference frame at the location of PS17cke (Figure \ref{fig:makeSubtractedFrameImage}). 
The flux corresponds to a $V$ band magnitude of $22.86 \pm 0.18$, measured via PSF fit. 
As with the Pan-STARRS1 subtractions, the high surface brightness of the host leaves subtraction residuals across the host and the detection of this positive flux corresponds to a marginal detection.
Either this is real positive flux, or given the errors, it corresponds to a 5$\sigma$ upper limit.
Either way, it confirms that the object faded rapidly and we consider both options in Section \ref{sec:ps17cke:disc}. 

We have also forced photometry in all difference images made by the ATLAS survey \citep{Tonry2018} to check for any outbursting signature at this sky position. 
As described by \citet{Smith2020}, ATLAS typically observes with a sequence of $4\times30$\,s exposures every two nights, with gaps due to weather, with difference images of detections of positive sources being carried out automatically.
We forced photometric PSF measurements at the the position of PS17cke and found no significant flux in the history of ATLAS imaging.
The fluxes relevant to the period around the PS17cke discovery epoch are reported in Table \ref{tab:my_label}, along with corresponding 5$\sigma$ upper limits for magnitudes.
We note as well that there appears to be no significant detections made by other groups such as ASAS-SN \citep{Shappee2014} or ZTF \citep{Bellm2019} at the location of PS17cke.

\begin{figure}
    \centering
    \includegraphics[width=\linewidth]{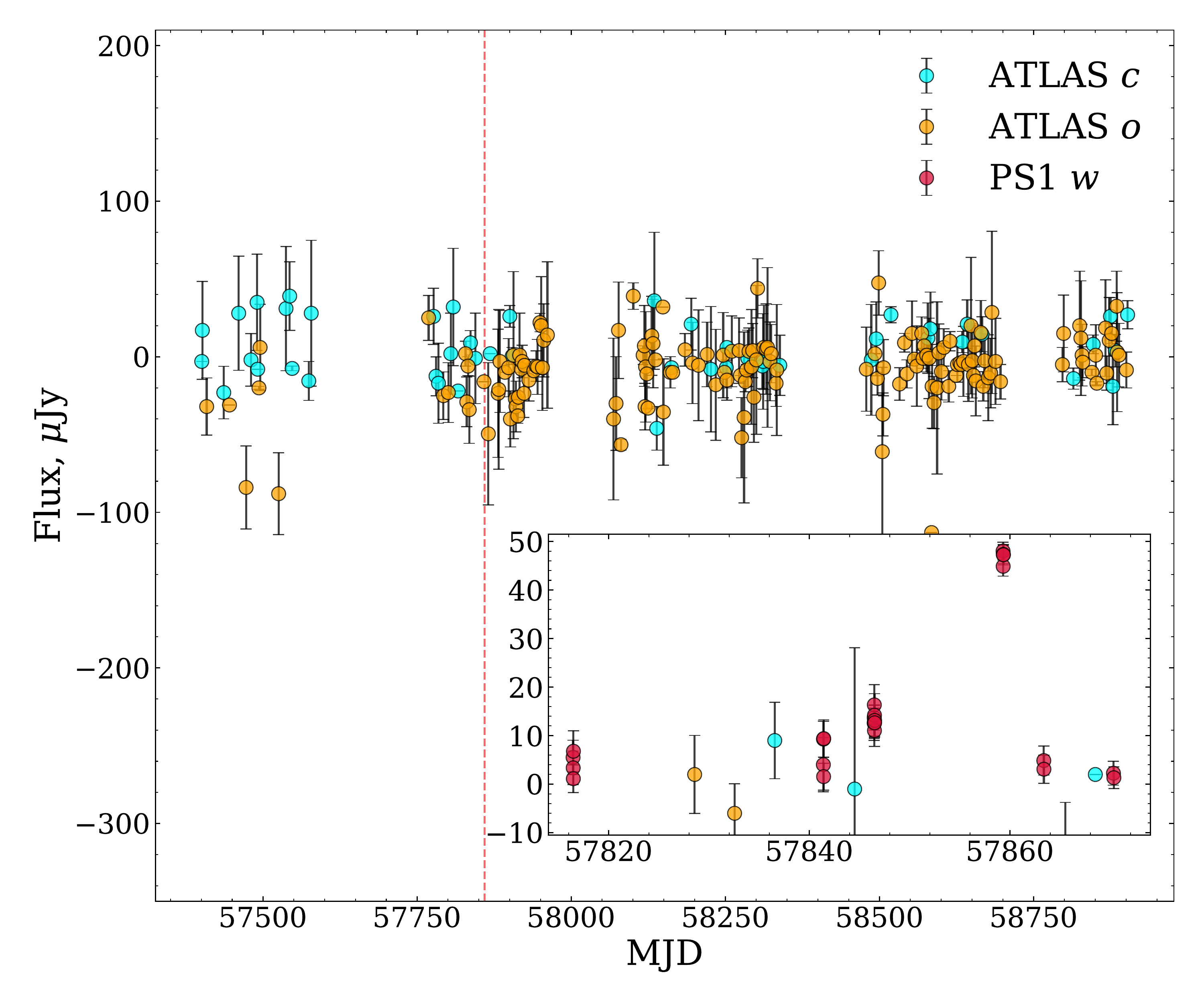}
    \caption{The ATLAS flux lightcurve for the entirety of its duration of operation forced at the location of PS17cke. The red dashed line denotes the Pan-STARRS discovery epoch of PS17cke. We show as well in the inset plot our PS1 forced photometry of PS17cke spanning the date range of MJD 57800 to MJD 57870 overlaid on the ATLAS data.
    }
    \label{fig:atlas_lc_ps17cke_snr5}
\end{figure}

\begin{figure}
    \centering
    \includegraphics[width=\linewidth]{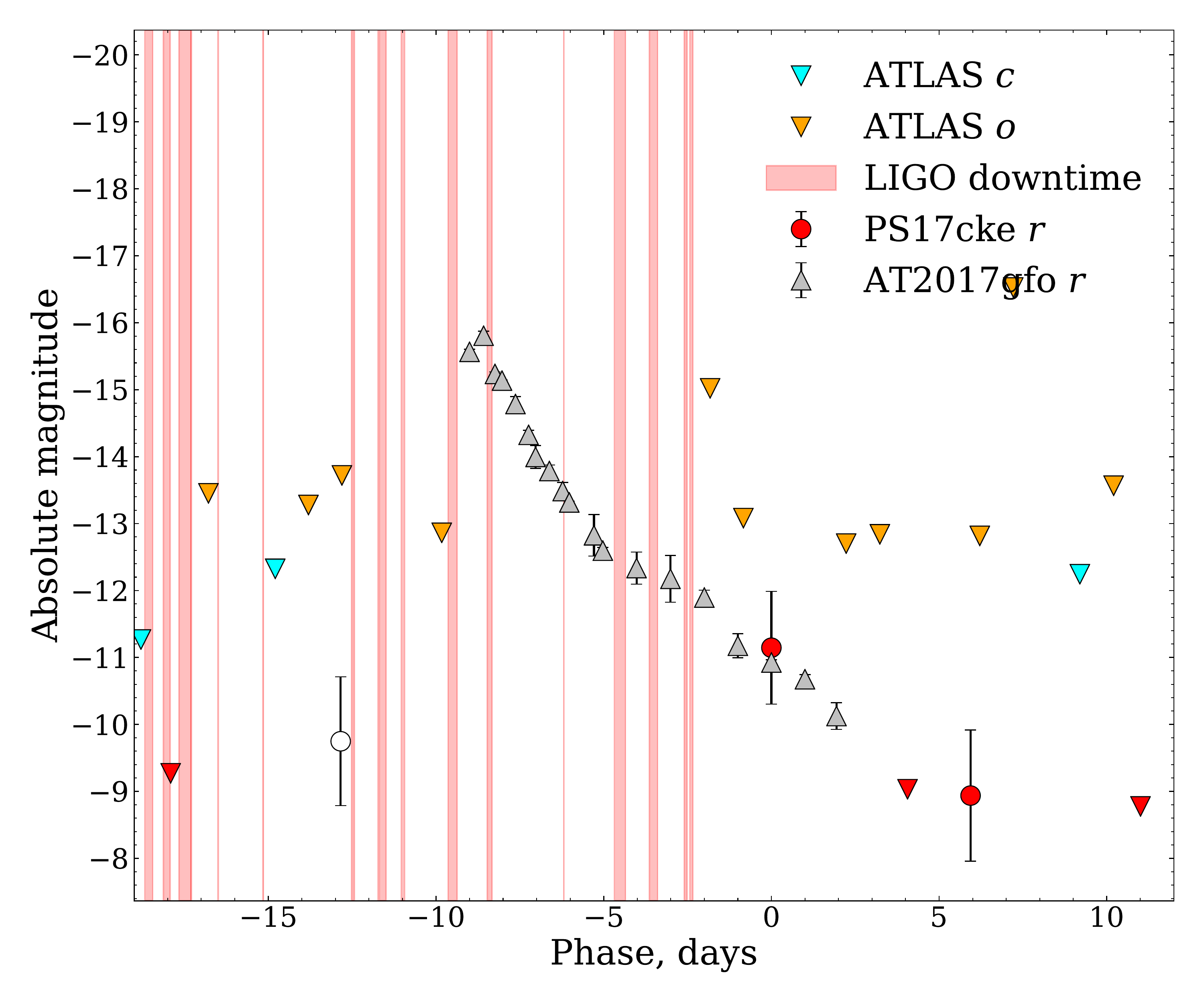}
    \caption{The \wps\ absolute lightcurve of PS17cke converted to Sloan-$r$ \citep{Tonry2012} with the discovery exposures and surrounding non-detections averaged per observing epoch. We note the origin of each data point - the +0.000 day point is the discovery detection with Pan-STARRS1, the -12.802 day point (shown as an open circle due to its uncertain nature) is the result of net flux being found in our Pan-STARRS forced photometry and the +5.950 day point is from net flux in the NTT:EFOSC2 acquisition frames for the attempted classification spectrum. Overlaid is the $r$ band lightcurve of AT2017gfo shifted in time to demonstrate the plausibility that a fast fading event such as a kilonova could have occurred in the period of time between the most recent observations before and after discovery, as well as upper limits from ATLAS as these are closer in time than for PS1. The red bands denotes the regions of time where neither LIGO detector was active in the days preceding discovery of PS17cke.
    }
    \label{fig:17cke_lightcure}
\end{figure}


\section{Discussion}
\label{sec:discussion}

\subsection{Interpretation of PS15cey}
\label{sec:ps15cey:disc}

The secure redshift measurement of the host of PS15cey puts it significantly beyond the 200\,Mpc distance limit that we originally imposed for the search for kilonovae (Smartt et al. 2020, in prep).
With the most recent non-detections occurring approximately 50 days before and after discovery, we note that it is possible for a Type II SN of some description to have occurred in this time, which would otherwise be a major contaminant in the PSSK.
Our temperature estimate for this object disfavours a Type II SN progenitor however, though we note it would be more consistent with a tidal disruption scenario \citep{Jiang2016}.
Tidal disruption can be ruled out, however, due to the offset from the host visible in Figure \ref{fig:ps15cey_triplet}.
As such, we regard PS15cey as a fast declining transient about which we will explore various kilonova models and also compare it to the fast transient AT2018kzr \citep{McBrien2019,Gillanders2020} in this section.
The peak magnitude of \rps$=-18.40$ puts PS15cey 2 magnitudes brighter than AT2017gfo at peak \cite[see the early photometry of][]{Arcavi2017,Coulter2017,Cowperthwaite2017,Evans2017,Smartt2017,Valenti2017}. 
To determine if this luminosity is plausible, we investigate models of black hole and neutron star mergers \citep[BH-NS;][]{Barbieri2020BHNS} and luminous kilonova (NS-NS) models in the vein of those presented by \citet{Barbieri2020gw190425}, which employ a `stiff' equation of state.

In the case of a NS-NS merger, high luminosity is achieved when the amount of dynamical ejecta is maximised and/or an accretion disk is produced.
A larger accretion disk size and total ejecta mass will also enhance the luminosity of any kilonova-type transient.
In the models of \citet{Barbieri2020gw190425}, the combination of neutron star masses that will achieve maximal luminosity occurs for a primary mass of 2.42\msun\ and secondary of 1.00\msun.
These are at the limits of what is permitted with one of the `stiffest' equations of state \citep[DD2, as explained by][]{Hempel2010,Typel2010}.
DD2 is consistent with both GW170817 \citep{LigoVirgo2019}, GW190425 \citep{LigoVirgo2020} and an AT2017gfo type kilonova in the framework of \citet{Barbieri2020gw190425}.
This system provides a highly deformable secondary which is disrupted by strong tidal forces from the primary.
However even these `bright' models do not produce a luminous enough event to describe PS15cey, typically peaking at luminosities which are $1.5 - 2$ magnitudes fainter than the peak of PS15cey.

In the case of a BH-NS merger, a higher ejecta mass can be obtained by increasing the black hole spin as this moves the innermost stable circular orbit of the binary closer to the black hole \citep{Barbieri2019}.
For a maximally spinning black hole ($\chi_{\mathrm{BH}} = 0.99$), using the same equation of state as in the NS-NS models, the maximum dynamical ejecta, approximately 0.24\msun, is achieved for a black hole mass of $\mathrm{m_{BH}} = 11.68$\msun\ and neutron star mass of $\mathrm{m_{NS}} = 1.00$\msun.
This configuration favours strong tidal disruption of the neutron star, giving rise to a large dynamical ejecta mass, as this material remains gravitationally unbound.
A different combination of component masses can maximise the mass of the accretion disk.
This is achieved with a system with  $\mathrm{m_{BH}} = 2.43$\msun\ and $\mathrm{m_{NS}} = 1.59$\msun, as it has a smaller innermost stable circular orbit and achieves an accretion disk mass of approximately 0.54\msun.
Similarly, total ejecta mass (as opposed to total dynamical ejecta) is maximised for a combination of masses of $\mathrm{m_{BH}} = 2.43$\msun\ and $\mathrm{m_{NS}} = 1.49$\msun, at approximately 0.55\msun.
Throughout these models, it is assumed that no mass gap exists between the black holes and neutron stars.
These BH-NS models are capable of producing lightcurves that peak around $-18$ magnitudes in both $r$ and $i$.

The model system which has ${m_{\rm BH}}=11.68$\msun, ${m_{\rm NS}}=1.00$\msun, a maximally spinning BH and a NS with the DD2 equation of state can produce sufficiently luminous emission to match  both the \rps\ and \ips\ data points.
We show these fits Figure \ref{fig:plotPS15ceyLightcurve}.
These fits hold for explosion epochs in the range of 0.8 to 1.2 days prior to discovery, with the specific profiles shown being for 0.8 days.
A large initial nuclear heating rate of $\varepsilon_{0} = 2 \times 10^{19}$\,erg\,s$^{-1}$\,g$^{-1}$ is also assumed.
Typical initial heating rates of r-process material varies in the region of approximately $10^{18} - 10^{19}$\,erg\,s$^{-1}$\,g$^{-1}$ \citep{Metzger2010} and follows a fitting function when its variation over time is considered, as outlined by \citet{Korobkin2012}.
This model bears many similarities to previous BH-NS models developed \citep[as in ][]{Barbieri2020gw190425}, but also has some important differences.
The main similarities are in the dynamical ejecta opacity ($\kappa_{\mathrm{dyn}} \simeq 20$\cmg) and opening angle $\theta_{\mathrm{dyn}} \gtrsim{} 17^{\circ}$.
The differences, however, span wind ejecta opacity $\kappa_{\mathrm{w}}$, wind and secular ejecta velocities, $v_{\mathrm{w}}$ and $v_{\mathrm{s}}$, and the fraction of the accretion disk flowing into both wind ejecta $\zeta_{\mathrm{w}}$ and secular ejecta $\zeta_{\mathrm{s}}$.
We have adopted values of $0.5$\cmg\ for $\kappa_{\mathrm{w}}$, which would be more typical of a NS-NS merger and 0.2\,c for both $v_{\mathrm{w}}$ and $v_{\mathrm{s}}$ which is twice what is normally used.
We have also employed 0.1 for $\zeta_{\mathrm{w}}$ with $\zeta_{\mathrm{s}}$ being unconstrained in the models.
A typical value for $\zeta_{\mathrm{w}}$ for a BH-NS merger is 0.01, so it would be somewhat surprising to see so much of the disk being directed into the wind ejecta.
Therefore a luminous BH-NS merger model is a possible description for the fast decline of PS15cey, but we acknowledge that the model parameters are at the extreme ends of the what are physically plausible.

At the time of discovery of PS15cey, LIGO would have been undergoing its first observing run, O1 \citep{Abbott2016BNS}. 
During this time, the detectors were sensitive to binary neutron star mergers with constituent masses $1.35 \pm 0.13$\msun\ to a volume-weighted average distance of ${\sim} 70$\,Mpc. For black holes of at least 5\msun, however, this distance stood at ${\sim} 110$\,Mpc. We discussed previously that at our chosen cosmology PS15cey is at a luminosity distance of 320\,Mpc, far beyond the LIGO O1 binary neutron star detection horizon, so would not have been detected if it were a BH-NS merger event.

From Figure \ref{fig:plotPS15ceyLightcurve}, we can see that PS15cey bears a strong resemblance to the fast transient AT2018kzr \citep{McBrien2019,Gillanders2020}.
AT2018kzr was notable for its rapid decline across optical bands at a rate similar to the kilonova AT2017gfo.
The interpretation of the photometry and spectroscopy of this object was that it was a potential merger between a white dwarf and neutron star (WD-NS).
AT2018kzr peaked in absolute magnitude at $-17.98$ in the $r$ band, comparable to PS15cey peaking at $-18.34$.
If PS15cey were to be similar to a WD-NS merger event such as AT2018kzr, we would expect it to display similar spectra and colours. 
AT2018kzr has colour \rps-\ips$=-0.18\pm0.02$ from the average of the synthetic photometry of the earliest two spectra and the earliest two epochs of multi-colour photometry presented by 
\citet{McBrien2019}.
Hence, the estimated \rps\ magnitude at this second epoch of PS15cey would be \rps=19.46, which implies a decline of 0.29\,mag\,day$^{-1}$. 
A similar conversion of the \ips\ magnitude to an \rps\ point using the 
colour of AT2017gfo at 1 to 2 days \citep[$r-i=0.07$ after extinction correction]{Smartt2017} implies an equivalent \rps\ band magnitude of 19.57 (for the second epoch), and an \rps\ decline rate of 0.40\,mag\,day$^{-1}$.
These values of $0.3 - 0.4$\,mag\,day$^{-1}$ imply a slightly slower decline rate to either an AT2017gfo-like kilonova or an AT2018kzr event, the latter of which declined at $0.48 \pm 0.03$\,mag\,day$^{-1}$ in $r$.
However the decline rate is more similar to these fast transients than a typical supernova.
It leaves open the possibility that PS15cey is an extreme type of kilonova, such as those resulting from BH-NS mergers as illustrated with the model comparison. 
Further searches for luminous, fast declining objects will be important in future surveys. 

\subsection{Interpretation of PS17cke}
\label{sec:ps17cke:disc}

The nature of PS17cke is dependent on the reality of the forced flux measurement on the night MJD 57846.506 (12.8 days before discovery) being real or not. 
As discussed in Section \ref{sec:ps17cke}, in each of the $8\times45$\,s images on that night, there is excess flux but each are $\lesssim{}5\sigma$.
Combining the flux measurements formally leads to a ${\sim}6\sigma$ detection above the zero level, but since there seems to be a baseline residual flux of around 6\,$\mathrm{\mu}$Jy, this excess is only 3$\sigma$ above that.
The forced photometry is shown in the inset panel of Figure\,\ref{fig:atlas_lc_ps17cke_snr5} where the level of the significance can be judged.

\subsection{PS17cke as a kilonova}  
\label{sec:17ckeKN}

We first assume that this flux on MJD 57846 is not significant and is due to imperfect subtractions at this position, due to the large surface brightness fluctuations.
From Figure \ref{fig:17cke_lightcure}, we can see that in the period of time allowed between that most recent Pan-STARRS1 non-detections before discovery (MJD 57846) and after discovery (MJD 57863) that it would be possible for an AT2017gfo-like event to occur and subsequently fade from the view of Pan-STARRS1.
In Figure \ref{fig:17cke_lightcure}, we have shifted the $r$ band lightcurve of AT2017gfo back in time by 9 days from maximum and note the decline of PS17cke from discovery to the next subsequent non-detection is of comparable rapidity.
Given this, it opens the possibility that if PS17cke were a genuine kilonova, it may have been detected after maximum light and on decline.
The decline of PS17cke from detection to below the detection limit occurs at a rate of approximately $0.523$\magday, which is indeed similar to the decline rate of AT2017gfo.
If it were similar to AT2017gfo, that would imply a merger date around MJD $57850.3 \pm 0.5$, 9 days prior to our PS1 discovery.
At this epoch, LIGO was observing with both the Livingston (L1) and Hanford (H1) units\footnote{\url{https://www.gw-openscience.org/detectorstatus/}}.
The combined uptime of both detectors achieved approximately 65\% completeness of the day to a depth of near 70\Mpc\ \citep{LigoVirgo2018}.
Unfortunately, without 100\% uptime of at least one detector, it is possible that any kilonova behind PS17cke could have occurred without GW detection.  
The data point from the NTT imaging on MJD 57865.258 (+5.950 days) suffers from the same uncertainty after image subtraction.
We do detect excess flux at the position of PS17cke, but at about 3$\sigma$ significance.
However, whether this is a detection of faint flux or an upper limit is not critical to the kilonova interpretation, since the level of the flux, or inferred limit, support the conclusion of a rapid fade within $4-6$ days.
Although we have only one secure data point, it is clear that this rapid fade is compatible with an AT2017gfo type kilonova. 

However, if the forced photometry flux on $-12.8$ days is real and secure, and the errors are reflective of a real detection, then the lightcurve is not compatible with AT2017gfo, as is clear from Figure \ref{fig:17cke_lightcure}.
There is no way to shift the lightcurve to match any part of AT2017gfo.
We may more quantitatively explore this by considering comparisons to the diverse array of kilonova lightcurve models that have been developed by several groups in recent years.
These account for different binary combinations, ejecta parameters and heating rates, so offer several plausible profiles to compare to.
We have chosen a number of these to see which provide a favourable comparison to the lightcurve of PS17cke, beyond the simple comparison to AT2017gfo in Figure \ref{fig:17cke_lightcure} and to determine if a kilonova interpretation is physically plausible if the first detection is real. 

\begin{figure}
    \centering
    \includegraphics[width=\linewidth]{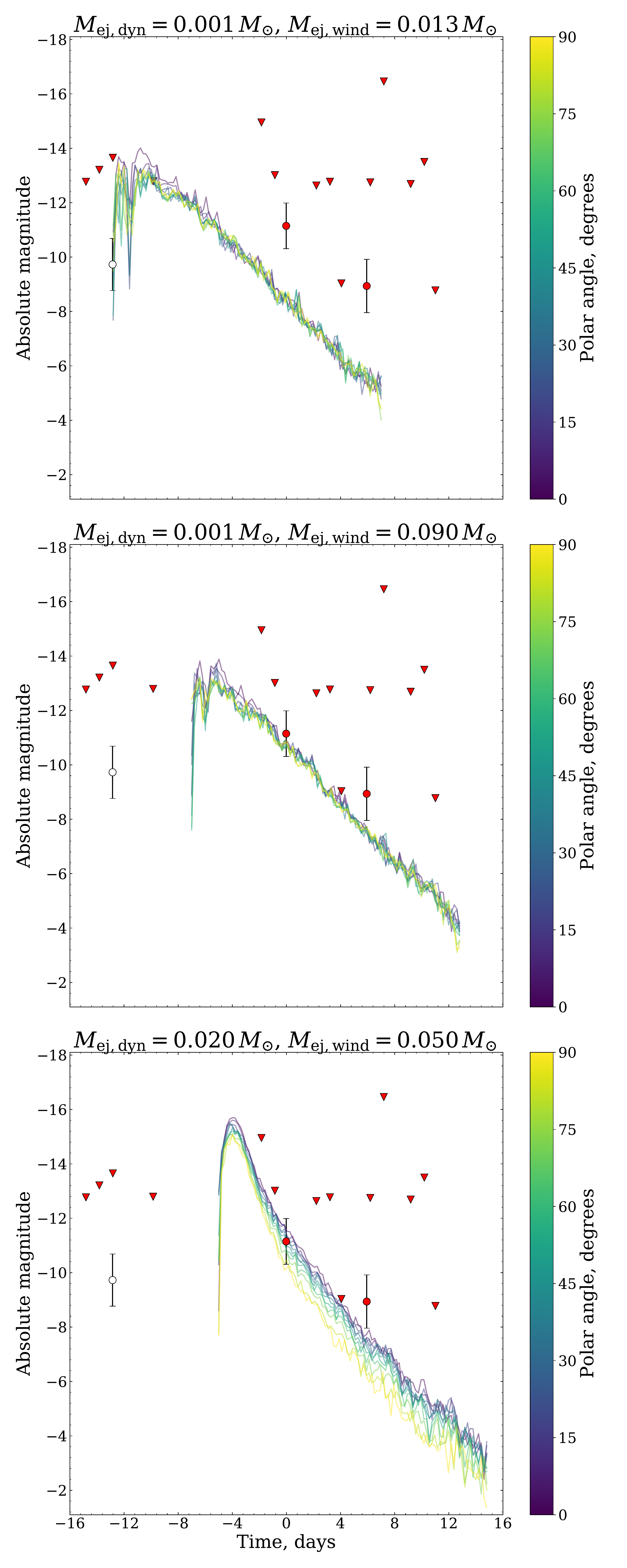}
    \caption{A comparison of the lightcurve of PS17cke with several models from the grid presented by \citet{Dietrich2020}. \textbf{Top}: An attempt to fit all three data points, assuming the earliest detection equates to the merger time. The best fit viewing angle of this plot is $15\degree$. \textbf{Middle}: An attempt to fit the final two data points in time using a merger time equal of 7 days prior to discovery. The best fit viewing angle of this plot is $60\degree$. \textbf{Bottom}: An attempt to fit the final two data points in time using a merger time equal of 5 days prior to discovery. The best fit viewing angle of this plot is $60\degree$.
    }
    \label{fig:plotCompareBullaModels}
\end{figure}

Our first attempt to model the lightcurve of PS17cke was achieved using the POlarization Spectral Synthesis In Supernovae code, \texttt{POSSIS}, presented by \citet{Bulla2019}. \texttt{POSSIS} is a three dimensional Monte Carlo radiation transfer code capable of synthesising flux and polarisation spectra of supernovae and kilonovae, as well as lightcurves for both, by treating opacity and several ejecta parameters as time-dependent properties.
The specific grid of models we are comparing to was presented recently by \citet{Dietrich2020} which give dynamical ejecta mass $M_{\mathrm{ej,dyn}}$, post-merger wind ejecta mass $M_{\mathrm{ej,wind}}$, half-opening angle of the lanthanide-rich region $\phi$ and viewing angle $\theta_{\mathrm{obs}}$ (given as $\cos \theta_{\mathrm{obs}}$) as free parameters.
We show some of our formal `best' fits to the $r$ band PS17cke data in Figure \ref{fig:plotCompareBullaModels}.

The top panel of Figure \ref{fig:plotCompareBullaModels} is an attempt to fit all three data points on the lightcurve assuming that the merger time corresponds to the date of the earliest detection.
It appears impossible to obtain a fit to all three data points, if the forced photometry flux at -12.8 days is real. 
The fits shown give ejecta masses that tend to the most extreme values allowed on the grid with $M_{\mathrm{ej,dyn}} = 0.001$\msun, the minimal allowed mass, and $M_{\mathrm{ej,wind}} = 0.13$\msun, the maximum allowed mass. 
The middle and bottom panels of Figure \ref{fig:plotCompareBullaModels} both discount this first data point and fit only the second and third.
They do so for different assumed merger times of 7 days prior to discovery in the middle panel and 5 days prior to discovery in the bottom panel.
The best fit parameters of these fits are $M_{\mathrm{ej,dyn}} = 0.001$\msun, $M_{\mathrm{ej,wind}} = 0.09$\msun, $\phi = 60^{\circ}$ and $\cos \theta_{\mathrm{obs}} = 0.7$ ($\theta_{\mathrm{obs}} = 45.57^{\circ}$) in the middle panel, and $M_{\mathrm{ej,dyn}} = 0.02$\msun, $M_{\mathrm{ej,wind}} = 0.05$\msun, $\phi = 15^{\circ}$ and $\cos \theta_{\mathrm{obs}} = 0.3$ ($\theta_{\mathrm{obs}} = 72.54^{\circ}$).
Both fits do adequately describe the data, but constraining the parameters any further than this requires a more precise knowledge of the merger epoch.
We show as well in these panels the combined ATLAS and PS1 upper limits from epochs surrounding the discovery of PS17cke, corrected from their native filters to $r$ by assuming a similar SED to AT2017gfo, but do not observe any discrepancies between the model fits and the upper limits.

In the case of the first model, which assumes the merger time was the same as our earliest photometric data point on MJD 57846.506 (2017-04-03 12:08:38 UTC), we note that both L1 and H1 were observing to depths of $70 - 80$\Mpc, though given the quality of the fit in the top panel of Figure \ref{fig:plotCompareBullaModels}, we do not believe that this is representative of a true merger epoch.
The second model, however, uses a merger time of 7 days prior to discovery (MJD 57852).
On this epoch, the combined detector uptime reached 100\% to similar distances as before.
Likewise, the third model used a merger time of 5 days prior to discovery (MJD 57854) where coverage from both LIGO units was more sparse.
We estimate the combined detector completeness was approximately 63\% on this epoch.

The recently published models of \citet{Kawaguchi2020} provide another suitable point of comparison.
These models, based on those of \citet{Tanaka2013}, also come from a radiative transfer code with photon transfer simulated by Monte Carlo calculations for various profiles of ejecta density, velocity and elemental abundances.
Of particular note among the \citet{Kawaguchi2020} models are those of the fiducial case of a binary neutron star merger that undergoes a prompt collapse to a black hole and a scenario in which the binary neutron star merger begets a long-lived supermassive neutron star (SMNS) remnant.
The chosen models, per Table 1 of \citet{Kawaguchi2020}, are \texttt{BH\_PM0.001}, \texttt{BH\_PM0.01}, \texttt{SMNS\_DYN0.003} and \texttt{SMNS\_DYN0.01}. \texttt{BH\_PM0.001} and \texttt{BH\_PM0.01} describe the prompt collapse of the merged neutron stars for lanthanide-rich post-merger ejecta masses of 0.001\msun\ and 0.01\msun\ respectively, though both have a dynamical ejecta mass of 0.001\msun.
On the other hand, \texttt{SMNS\_DYN0.003} and \texttt{SMNS\_DYN0.01} describe the formation of a long-lived SMNS following the merger which accelerates the post-merger ejecta to higher velocities than in the previous models.
For these, a common post-merger ejecta mass of 0.05\msun\ is adopted while dynamical ejecta masses of 0.003\msun\ and 0.01\msun\ respectively are employed.

\begin{figure*}
    \centering
    \includegraphics[width=\linewidth]{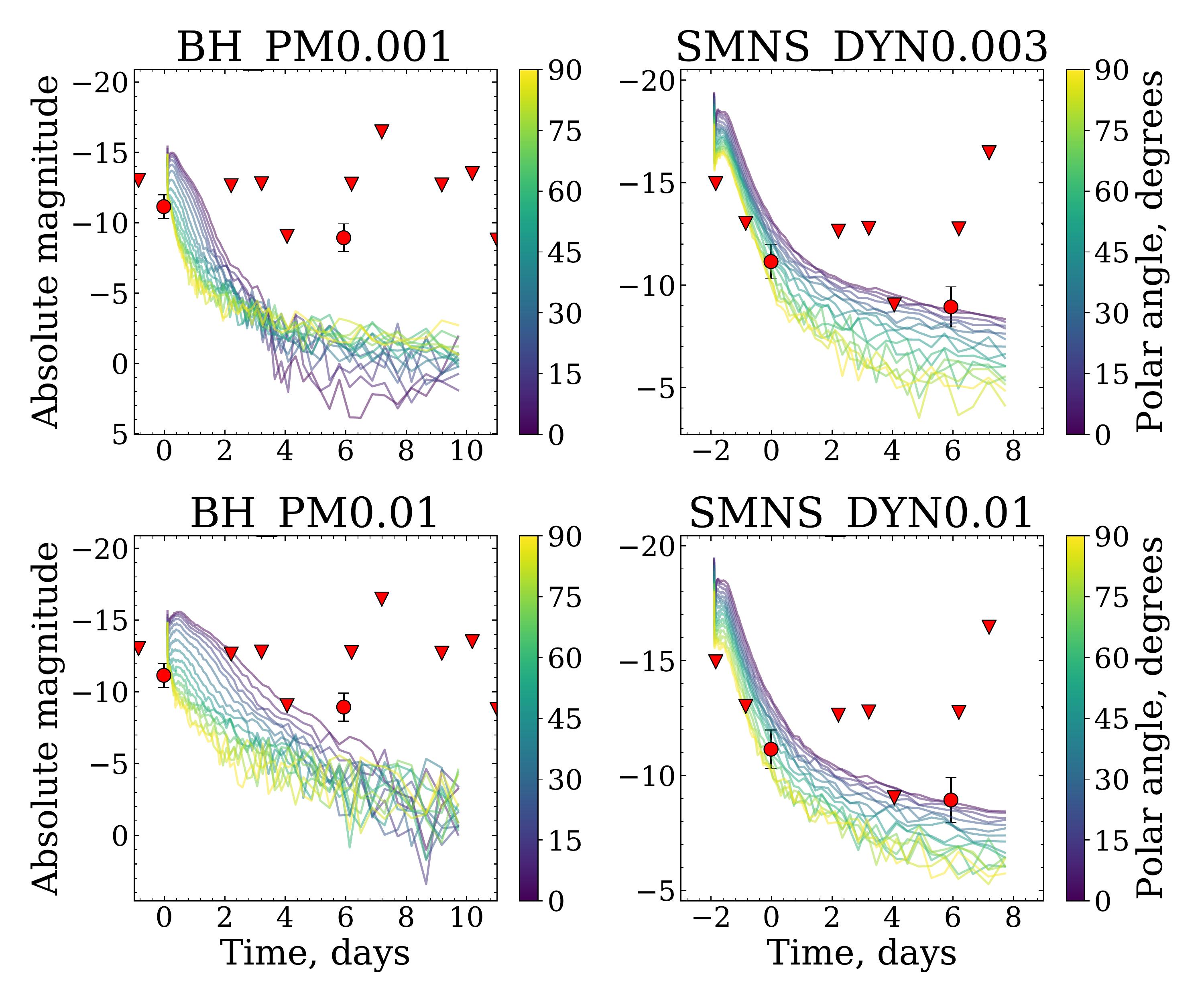}
    \caption{A comparison of the PS17cke lightcurve with several models presented by \citet{Kawaguchi2020}. The chosen models include ejecta profiles that describe scenarios of prompt collapse to a black hole and the formation of a long-lived supermassive neutron star (SMNS), with various combinations of post-merger and dynamical ejecta masses. \textbf{Top left}: A prompt collapse model with post-merger ejecta mass 0.001\msun. \textbf{Bottom left}: A prompt collapse model with post-merger ejecta mass 0.01\msun. \textbf{Top right}: A SMNS model with dynamical ejecta mass 0.003\msun. \textbf{Bottom right}: A SMNS model with dynamical ejecta mass 0.01.
    }
    \label{fig:plotCompareKawaguchiModels}
\end{figure*}

\citet{Kawaguchi2020} present simulated lightcurves of these merger scenarios for $grizJHK$ and we show the comparison of the decline of PS17cke to the $r$ band models in Figure \ref{fig:plotCompareKawaguchiModels} for a distribution of viewing angles on the ejecta over $0^{\circ} < \theta < 90^{\circ}$.
As with the POSSIS models, we include the ATLAS and PS1 upper limit data, converted to $r$, when interpreting the model comparisons.
There are no models that can fit the data if we assume that the first point at -12.8 days is real and a reliable measurement.
If it is treated as a non-detection and upper limit then at least three of the models can reasonably reproduce the data, though with some variation in the predicted merger time between them.
In the case of the prompt collapse models, they are typified by low ejecta masses which is supported by numerical simulations performed by \citet{Hotokezaka2013}.
The ejecta becomes suppressed by the prompt collapse, which in turn contributes to a lower luminosity lightcurve.
We see this in the left panels of Figure \ref{fig:plotCompareKawaguchiModels}, with all lightcurves being fainter than at least -15.5 magnitudes for both model sets. 
Of these models, if we were to assume the merger time is equivalent to or in the region of the discovery epoch of PS17cke, we would observe that \texttt{BH\_PM0.01} provides a better fit than \texttt{BH\_PM0.001}, with both being broadly consistent with the limits imposed by ATLAS and PS1.
At the same time, both the \texttt{SMNS\_DYN0.003} and \texttt{SMNS\_DYN0.01} models provide good fits to the data if the merger time is taken to be 2 days before discovery (MJD 57857).
We must note here that due to uncertainty in the opacity values used to produce these models, it is possible their brightness at epochs $\lesssim 1$ day is overestimated.
Hence, we do not attribute a large significance to the coincidence of \texttt{BH\_PM0.01} with PS17cke or the discrepancy of the ATLAS and PS1 upper limits with \texttt{SMNS\_DYN0.003} and \texttt{SMNS\_DYN0.01} that can been seen in the right panels of Figure \ref{fig:plotCompareKawaguchiModels}.

On the epoch of discovery the combined detector uptime was approximately 98\% to a typical distance of 60\Mpc.
Two days prior to this, the merger epoch inferred by the SMNS model comparisons, we estimate the combined detector uptime to be approximately 99\% to a similar distance.

\subsection{PS17cke as a luminous blue variable outburst} 

If the first data point at -12.8 days is real, then the analysis in Section \ref{sec:17ckeKN} indicates that it is difficult to interpret the transient as a kilonova. 
Another possible explanation for the origin of such an intrinsically faint event is in an outburst from an unstable massive star such as a luminous blue variable \citep[LBV,][]{Humphreys1994}.
LBVs are often discovered in nearby ($d < 25$\Mpc) galaxies at absolute magnitudes $-11 > M > -13$ which is consistent with this object. 
A comparison of the lightcurve of PS17cke to previously discovered objects with notable outbursts offers some insight to this object being a LBV.
SN2009ip was first discovered on 2009-08-26 \citep{Maza2009} and has been observed for several years since its discovery.
It notably showed rebrightening several years later \citep{Drake2012} and is now known to be the result of an outbursting LBV in NGC 7259.
Observations presented by \citet{Pastorello2013} in the years since its discovery provide an extensive $UBVRI$ lightcurve for comparison.
For PS17cke to fade below the most recent non-detection following discovery it must decline by 0.523\,mag\,day$^{-1}$.
By simply measuring the rise and decline between successive points in the $R$ band lightcurve of SN2009ip we can identify at least five regions where the decline exceeds this limit for PS17cke.
Similarly, we can compare PS17cke to the LBV in NGC 3432, formally known as SN2000ch \citep{Pastorello2010LBV}. 
There are at least two regions of the $V$ band lightcurve presented by \citet{Pastorello2010LBV} which exceed the decline rate limit.
While this is not conclusive proof that PS17cke is a LBV outburst, it indicates it is a plausible explanation. 

Although numerous LBVs have been discovered which have exhibited repeat outbursts, perhaps most famously $\eta$ Carinae \citep{Humphreys1999} but also the events discussed by \citet{Maund2006} and \citet{Smith2011} as well as \citet{Pastorello2010LBV,Pastorello2013}, it is unclear if this is a ubiquitous trait of this class of objects.
Of course, supposing PS17cke is a LBV, the detection of a repeat outburst would immediately disqualify it from consideration as a kilonova candidate.
As discussed in previous sections, we have forced photometry throughout the history of Pan-STARRS observations between MJD 57805 and MJD 58879 and find no significant flux excess above 21$^{\mathrm{st}}$ magnitude.
For more complete temporal coverage and a search for activity at the position of PS17cke, we 
have analysed all ATLAS data at this position \citep{Tonry2018,Smith2020}.
We have generated forced photometry at the location of PS17cke from the ATLAS images and manually inspected all exposures for the duration of its survey.
We note that there are significant detections (>5$\sigma$) at several epochs including MJD 57893, MJD 57904 and MJD 58545.
An inspection of the exposures show that the majority of these are failed subtractions resulting in excess flux from the host contaminating the forced photometry measurement or other detector artefacts and we rule them out as real detections. 
We have measured our archival forced photometry at the location of PS17cke in ATLAS in units of flux density (microJansky) and plotted this in Figure \ref{fig:atlas_lc_ps17cke_snr5}.
That we do not detect PS17cke is perhaps unsurprising as the discovery magnitude in PS1 was  \wps=$19.69 \pm 0.04$
For ATLAS, this is below the detection limit, so we would require an even brighter outburst to occur for ATLAS to detect it.

What is necessary to conclusively identify whether PS17cke is a LBV is deeper imaging data which shows a luminous point source at its location.
We have inspected the deeper imaging data from NTT:EFOSC2, (totalling 1200\,s in the $V$ band) from MJD 58872.319 (2020-01-24 07:39:42 UTC). 
The image quality of the combined frame was 0\farcs69 but this was not deep enough or high enough in resolution to determine if a stellar point source (or compact stellar cluster) exists at the position of PS17cke.
There is no clear detection of a resolved source, but there is some extended emission at the position.
To conclusively distinguish between a LBV type outburst and a kilonova candidate will require Hubble Space Telescope imaging, both in the optical and NIR.
This would determine if there were a point source to absolute magnitudes of $M \sim -5$ and hence if it is a stellar outburst.
If there was no detection of a point source then a faint transient such as a kilonova is a candidate. 

\subsection{PS17cke as a faint transient} 

We have discussed two promising explanations of the nature of PS17cke - those of a kilonova borne of either a NS-NS or BH-NS merger and the outburst of a LBV star.
In recent years, several intrinsically faint and fast fading transients have been discovered, themselves explainable through unique explosion and merger scenarios.
We outline a few in this section and briefly discuss the plausibility of a comparison between them and PS17cke.

\subsubsection{Fast evolving luminous transients}

Fast evolving luminous transients (FELTs) are a category of many classes of transient phenomena characterised by a rise to maximum of ${<}10$ days and subsequent fade from view in ${<}30$ days.
Numerous FELTs have been detected over the years, with prominent examples being those of KSN2015K \citep{Rest2018}, iPTF14gqr \citep{De2018} and iPTF16asu \citep{Whitesides2017}, as well as several of the objects discussed by \citet{Drout2014}.
While these objects do decline rapidly \citep[0.21\,mag\,day$^{-1}$ in $g$ in the case of iPTF14gqr,][]{De2018}, they are often too bright at maximum light to be compared to a kilonova progenitor, or are just too long lived.
In this paper, we presented model comparisons of NS-NS merger scenarios that spanned a peak absolute magnitude range of $-10 > M_{r} > -15$ and BH-NS merger scenarios that can reach $M_{r} \simeq -18$.
For context, the brightest of the above objects is iPTF16asu, peaking at $-20.4$ in $g$ \citep{Whitesides2017}.
An object as bright as these occuring in the host of PS17cke would have been detected at an apparent magnitude of $21 > m_{r} > 16$, well above the detection limit of PS1 and, for many objects, above that of ATLAS as well.
With regards to our ATLAS limits, we note the most constraining limits are between -9.828 days at $o>17.95$ and -0.835 days at $o > 17.73$ (there is an additional limit between these at -1.827 days, but this is shallower at $o > 15.79$).
Any FELT that could describe PS17cke would need to essentially fit into this ${\sim}9$ day period without being discrepant with our PS1 discovery point, but none of the outlined objects above are capable of this.

\subsubsection{Fast, blue optical transients}

Fast, blue optical transients (FBOTs) evolve on similar timescales to FELTs but also exhibit a strong blue colour ($g - r < -0.2$) and blackbody temperatures of $10,000$ to $30,000$\,K \citep[for a review of this category of objects, we refer the reader to that of ][]{Inserra2019}.
Objects that can be considered FELTs, such as those discussed in the previous section, are often considered to be FBOTs as well.
Some standout examples of FBOTs, particularly because of their blue colour, include SN2018gep \citep{Ho2019}, AT2018cow \citep{Prentice2018,Perley2019} and SN2019bkc \citep{Chen2019,Prentice2020}.
We encounter similar problems as with FELTs when comparing FBOTs to PS17cke in that they are often too bright or their evolution is not as rapid as would be necessary to describe PS17cke.
FBOTs have the additional diagnostic of colour and temperature information, but unfortunately this is lacking for PS17cke.
Still, we believe that if PS17cke were an FBOT that it should have been observed by ATLAS at at least one epoch.
As with FELTs, any FBOT that could describe PS17cke would need to be shorter lived than ${\sim}9$ day else it would have been observed by ATLAS.
The objects discussed here represent some of the fastest evolving events currently known, but it appears that the only AT2017gfo is capable of this feat.

\subsubsection{Type Iax supernovae}

Type Iax supernovae (SNe Iax) are a subclass of normal Type Ia supernovae characterised by typically lower luminosities and lower ejecta velocities.
Their classification name is derived from the prototypical SN Iax event, SN2002cx \citep{Li2003,Foley2013}.
For a review of this type of object, we refer the reader to that of \citet{Jha2017}.
Several objects belonging to this subtype have been discovered over the years that could provide a worthy comparison to PS17cke, including SN2008ha \citep{Valenti2009,Foley2009}, SN2010ae \citep{Stritzinger2014} and SN2019gsc \citep{Srivastav2020}.
Respectively, these objects peaked at $-14.2$ in $V$, $-15.3$ in $V$ and $-14.0$ in $r$.
We note that these are somewhat brighter than PS17cke, but is plausible given that it likely would have been discovered later in its evolution.
These SN Iax events display varying declines rates as well.
SN2008ha was quoted as declining fastest in bluer bands with $\Delta m_{15}(B) = 2.03 \pm 0.20$\,mag while for SN2010ae this stands at $\Delta m_{15}(B) = 2.43 \pm 0.11$\,mag \citep{Stritzinger2014}.
SN2019gsc is quoted as declining at a rate $\Delta m_{15}(r) = 0.91 \pm 0.10$\,mag, which gives an approximate nightly decline rate of ${\sim}0.06$\,mag\,day$^{-1}$.
Although similar in magnitude, these events prove to be too longer lived than PS17cke, which disfavours them for comparison.
It is difficult to compare further without multi-wavelength coverage or spectroscopic follow-up.

\subsubsection{WD-NS mergers}

We previously compared PS15cey to the suspected WD-NS merger AT2018kzr \citep{McBrien2019,Gillanders2020}, so for completeness shall do the same for PS17cke.
AT2018kzr was a significantly more luminous event than for what we observe in PS17cke, peaking at $-17.98$ in $r$, which places it in the category of a FELT.
Unlike other FELTs, however, AT2018kzr exhibited a strikingly rapid decline, approaching the rate at which AT2017gfo-like kilonovae decline, at $0.48 \pm 0.03$\magday\ in $r$.
We find this decline rate a much more favourable comparison than the previous objects considered.
The difference in peak magnitude can be explained if PS17cke is later in its evolution than our lightcurve suggests.
This is plausible, given the near 12 day gap in observations prior to discovery.
For reference, AT2018kzr had a rise time of within 3 days of its discovery, which, if a similar mechanism is responsible for PS17cke, we note is not discrepant with the data, even though it is not significantly constraining.
Similarly, as AT2018kzr approaches the rapid decline of AT2017gfo, it would be plausble that a WD-NS merger could have occurred in the time between the previously outlined constraining ATLAS limits and the PS1 detection of PS17cke.

\subsubsection{Luminous red novae}
\label{sec:lrne}

One possible interpretation of PS17cke is in the form of a massive stellar explosion such as a LBV, however, LBVs are not the only stellar explosions we may compare to.
Luminous red novae (LRNe) are another type of stellar explosion that result from the merging of two stars rather than as a consequence of instability and outburst from a single star.
The earliest example of a LRN is M85 OT2006-1 \citep{Kulkarni2007} which was discovered in the outskirts of M85 in the Virgo cluster, though its classification as a LRN is debated \citep{Pastorello2019}.
Since its discovery, numerous other LRN candidates have been identified, displaying similar characteristics, with magnitudes peaking in the gap between those of novae and supernovae.
Examples of these include AT2017jfs \citep{Pastorello2019at2017jfs} and AT2018hso \citep{Cai2019}.
A study of LRN performed by \citet{Pastorello2019}, which included six other LRN candidates beyond those mentioned here, found that these objects typically occupy magnitude space between $-13$ and $-15$ magnitudes and in many cases exhibit a secondary peak and plateau in their lightcurves.
While it is reasonable to assume that PS17cke did occupy this magnitude range, we do not observe a plateau for it or indeed any additional rebrightening of the lightcurve.
As such, we do not feel PS17cke is likely to be a LRN.

\subsubsection{Failed supernovae}

Lastly, we consider the possibility that PS17cke may be `failed supernova' - the result of a massive star collapsing to a BH without undergoing a SN explosion \citep{Heger2003}.
A survey for failed supernova candidates presented by \citet{Gerke2015} with the Large Binocular Telescope (LBT) yielded one event of note which \citet{Adams2017} refer to as N6946-BH1.
N6946-BH1 was a red supergiant (RSG) star in the galaxy NGC 6946 that experienced an optical outburst before subsequently fading from view \citep{Adams2017}.
The loss in brightness from outburst to fade spanned at least 5 magnitudes, though was not well sampled in $UBVR$ or $gri$ imaging from LBT or the Isaac Newton Telescope.
The temporal sampling of the NGC6946-BH1 candidate does not provide a decline rate on the few day timescale that we observe for PS17cke.
However, there is a physical reason to disfavour this scenario.
The death of a RSG in the mass range of $18 \lesssim \msol \lesssim 25$ has been predicted, through hydrodynamical simulations, to produce a short, optical transient (\logl\ $\simeq 10^{7}$ for approximately $3 - 10$ days) that gives rise to fainter (\logl\ $\simeq 10^{6}$), but longer lived (${\sim}1$ year) emission from the disruption of its hydrogen envelope following collapse of the helium core \citep{Lovegrove2013}.
The secondary emission is noted to be much redder, reminiscent of LRNe.
The \citet{Lovegrove2013} timescale of approximately 1 year is due to the dynamical timescale for a massive RSG.
The time for free fall collapse of a spherical body is $t_{\mathrm{dyn}} \sim \left( \frac{2 R^3}{GM} \right)^{\frac{1}{2}}$ giving a timescale of 0.5 years for a 20\msun\ RSG with $R \simeq 1000\rsun$.
Therefore, the rapid decline is not physically consistent with either the dynamical timescale of the star, nor the hydrodynamical calculation of \citet{Lovegrove2013}.


\section{Conclusions}
\label{sec:conclusions}

We have presented PS15cey and PS17cke (AT2017des) - two prospective kilonova candidates discovered as part of the Pan-STARRS Search for Kilonovae (PSSK).
PS15cey stood out for possessing a rapid decline approaching that of the kilonova AT2017gfo across two optical bands in Pan-STARRS.
Owing to uncertainty in available catalogue redshifts, we obtained spectroscopic observations of the host of PS15cey and measured a secure redshift, $z = 0.0717 \pm 0.0006$, which placed it outside the bounds of PSSK, at a luminosity distance of 320\Mpc. 
The redshift measurement also revealed that its peak absolute magnitude was at least two magnitudes brighter than AT2017gfo, which lead us to disfavour a NS-NS merger progenitor scenario for this object. 
The most promising attempts to model the available photometry were achieved with luminous kilonova models borne of BH-NS merger scenarios \citep{Barbieri2020gw190425}.
The models presented in this paper can successfully reproduce the peak magnitude of PS15cey and suggest a merger epoch within 0.8 to 1.2 days prior to our discovery.
Unfortunately, with the next observation of the field not occurring until 55 days after discovery, it is difficult to constrain the lightcurve fits further.
The specific models shown favour a maximally spinning black hole and fits can be produced to maximise dynamical ejecta and total ejecta mass for differing combinations of primary and secondary masses.
Comparisons to AT2018kzr also seem promising, both being of a similar peak magnitude and showing a similar decline. Without proper spectroscopic follow-up, however, it remains difficult to discern between these two possibilities and either are plausible. 

We conclude that our second candidate, PS17cke, is a plausible kilonova candidate.
With one secure detection and deep limits indicating a rapid fade, the decline rate is similar to AT2017gfo and a range of kilonova model lightcurves. 
We carried out forced photometry at the position of PS17cke in the history of both the Pan-STARRS and ATLAS surveys and find one marginal detection at -12.8 day before discovery.
We employed several checks to verify this including the construction of a customised deep reference stack in PS1 and creation of difference images using two different subtraction and analysis pipelines \citep{Rest2005,Magnier2016data}. 
The flux at this epoch can be attributed to residuals from the subtraction process, and it can not be confirmed as a genuine point source detection. 
Interpretation of the physical nature of PS17cke is dependent on whether this earliest point is real.
Assuming it is not, then our detections and limits allow a favourable comparison of the lightcurves of several NS-NS and BH-NS merger models, including those presented by \citet{Dietrich2020} and \citet{Kawaguchi2020}. 
These models match the measured decline rate and are consistent with the upper limits available. 
Depending on the model, we suspect the merger would have occurred between 2 and 7 days prior to discovery.
For the majority of this time, we note that LIGO was observing with at least one detector online, as is illustrated in Figure \ref{fig:17cke_lightcure}, making it highly likely that were a GW signal produced in concurrence with the optical signal of PS17cke, it would have been observed.
However there are periods during which neither LIGO detector was locked and working 
and it is possible that a merger occurred during one of these windows and was missed. 
If, however, this earliest data point is interpreted as a real detection, then a kilonova model is inadequate in describing the data and we consider a LBV outburst explanation more likely. 
Previously observed LBV outbursts, such as SN2000ch \citep{Pastorello2010LBV} and SN2009ip \citep{Maza2009,Pastorello2013}, do experience multiple periods of decline in their lightcurves comparable to PS17cke, but we only observe this outburst once for PS17cke in the history of PS1 and ATLAS survey data.
This is peculiar as we would expect multiple periods of outbursting were this a genuine LBV. 
We consider other faint and fast declining transients but find these even less likely to fit the observational constraints.
We propose that the two scenarios of a kilonova and a LBV outburst could be distinguished by deep Hubble Space Telescope imaging of the site.
The LBV scenario predicts that a luminous star must be visible at the position at $M\lesssim -5$, while the kilonova explanation would survive if there were no single massive star
at the location.
This would imply a kilonova explanation is plausible but not definitive. 
One final point we note is that the uncertainty in the explosion epoch of PS17cke, and PS15cey as well, renders the search for a high energy counterpart of the level of GRB170817A, rescaled to their respective distances moot, and that no high energy counterpart temporally or spatially coincident with either event has been recorded.

The kilonovae candidates presented in this paper represent examples of the kind of objects we are searching for in the Local Universe independent of GW triggers within the Pan-STARRS Search for Kilonovae \citep{Smartt2019AN}
and the ATLAS volume limited survey \citep[within 100\,Mpc]{Smith2020}. 
This may be a fruitful search for degenerate mergers in the gap between the LIGO-Virgo Consortium observing runs of O3 and O4 and may also discover
kilonovae beyond the horizon distances of the GW detectors or in periods when they
are not locked and observing.

\section*{Acknowledgements}
The authors would like to thank Russ Laher, Frank Masci and Adam Miller for checking the intermediate Palomar Transient Factory database for detections of PS15cey and Kyohei Kawaguchi for providing model data used in the comparisons for PS17cke.

SJS, KWS and DRY acknowledges funding from STFC Grant Ref: ST/P000312/1 and ST/S006109/1. 
CB acknowledges support from INFN, under the Virgo-Prometeo initiative.
MG is supported by the Polish NCN MAESTRO grant 2014/14/A/ST9/00121.
MF is supported by a Royal Society - Science Foundation Ireland University Research Fellowship.
MN is supported by a Royal Astronomical Society Research Fellowship.
KM acknowledges support from ERC Starting Grant grant no. 758638.
TWC acknowledges the EU Funding under Marie Sk\l{}odowska-Curie grant agreement no. 842471.
TW is funded by European Research Council grant 320360 and by European Commission grant 730980.
TMB was funded by the CONICYT PFCHA/DOCTORADOBECAS CHILE/2017-72180113.
FOE acknowledges support from the FONDECYT grant no. 1201223.

The discoveries from this program are a byproduct of the Pan-STARRS NEO survey observations. Operation of the Pan-STARRS1 and Pan-STARRS2 telescopes is primarily supported by the National Aeronautics and Space Administration under Grant No. NNX12AR65G and Grant No. NNX14AM74G issued through the SSO Near Earth Object Observations Program.

The Pan-STARRS1 Surveys (PS1) and the PS1 public science archive have been made possible through contributions by the Institute for Astronomy, the University of Hawaii, the Pan-STARRS Project Office, the Max-Planck Society and its participating institutes, the Max Planck Institute for Astronomy, Heidelberg and the Max Planck Institute for Extraterrestrial Physics, Garching, The Johns Hopkins University, Durham University, the University of Edinburgh, the Queen's University Belfast, the Harvard-Smithsonian Center for Astrophysics, the Las Cumbres Observatory Global Telescope Network Incorporated, the National Central University of Taiwan, the Space Telescope Science Institute, the National Aeronautics and Space Administration under Grant No. NNX08AR22G issued through the Planetary Science Division of the NASA Science Mission Directorate, the National Science Foundation Grant No. AST-1238877, the University of Maryland, Eotvos Lorand University (ELTE), the Los Alamos National Laboratory, and the Gordon and Betty Moore Foundation.

ePESSTO observations were obtained under ESO program ID 1103.D-0328 and 199.D-0143 (PI: Smartt).
ePESSTO+ observations were obtained under ESO program ID 1103.D-0328 (PI: Inserra).
SALT:RSS observations were obtained under program ID 2016-1-MLT-007 (PI: Jha).

This work has made use of the Asteroid Terrestrial-impact Last Alert System (ATLAS) project.
ATLAS is primarily funded to search for near earth asteroids through NASA grants NN12AR55G, 80NSSC18K0284, and 80NSSC18K1575; byproducts of the NEO search include images and catalogs from the survey area.
The ATLAS science products have been made possible through the contributions of the University of Hawaii Institute for Astronomy, the Queen's University Belfast, and the Space Telescope Science Institute.

\section*{Data availability}

The data underlying this article are available in the article and in its online supplementary material.

\bibliographystyle{mnras}
\bibliography{ref}

\end{document}